\documentclass[showpacs,preprintnumbers,amsmath,amssymb,aps,prd,preprint,nofootinbib]{revtex4}
\pdfoutput=1
\usepackage{graphicx}
\usepackage{amsmath,amssymb}
\usepackage{epstopdf}
\usepackage{verbatim}
\usepackage{mathdots}
\usepackage{color}%

\usepackage[colorlinks,  linkcolor=wine-stain,
urlcolor=blue,citecolor=blue]{hyperref}
\def\hhref#1{\href{http://arxiv.org/abs/#1}{arXiv:#1}} 

\definecolor{wine-stain}{rgb}{0.,0.,0.5}

\def\cI{{\mathcal I}}

\begin{document}

\title{Uniform WKB, Multi-instantons, and  Resurgent Trans-Series}

\author{Gerald V. Dunne}
\affiliation{Physics Department, University  of Connecticut, Storrs, CT, 06269}
\author{Mithat \"Unsal}
\affiliation{Department  of Physics,  North Carolina State University, Raleigh, NC, 27695}

\begin{abstract}

 We illustrate the physical significance and mathematical origin of resurgent trans-series expansions for energy eigenvalues in quantum mechanical problems with degenerate harmonic minima, 
by using the uniform WKB approach.  We provide evidence  that the perturbative expansion, combined with a global eigenvalue condition, contains all information needed to generate all orders of the non-perturbative multi-instanton expansion.   This provides a dramatic realization of the concept of  resurgence, whose structure is naturally encoded in the  resurgence triangle. We explain the relation between the uniform WKB approach, multi-instantons, and resurgence theory.  The essential idea applies to any perturbative expansion, and so is also relevant for quantum field theories with degenerate minima which can be continuously connected to quantum mechanical systems. 

\end{abstract}
\pacs{11.15.-q, 11.15.Kc ,11.15.Tk, 12.38.Aw, 12.38.Cy}
\keywords{Non-perturbative quantum field theory; renormalons; sigma models}
\maketitle

\tableofcontents

\section{Introduction}

\subsection{Why resurgent trans-series are important?}
In a large variety of quantum theoretical settings, it is  well known that 
 perturbative (P) and non-perturbative (NP) physics are closely  related \cite{Bender:1969si,Lipatov:1976ny,zinnbook,zinn-review}. 
In quantum mechanical  systems with degenerate (harmonic) minima, perturbation theory leads to divergent and non-alternating series \cite{Brezin:1977gk,Stone:1977au,Bogomolny:1980ur,Achuthan:1988wh}.\footnote{In this paper, we are concerned with quantum mechanical systems 
which only admit real instantons.  In more generic  cases where there are both real and complex saddles,  the connection between perturbation theory and non-perturbative saddles is more involved.   An example of this type of  more general problem is discussed  in \cite{bdu-inprep}.}
 This leads to two interrelated fundamental problems: 
 \begin{itemize}
 {\item[(i)] analysis of these divergent series (for example, by  Borel summation) leads to imaginary contributions to observables (such as energy) that must  be real;}
 {\item[ (ii)] this Borel summation procedure is ambiguous, with the ambiguity manifest in the sign of the imaginary non-perturbative contributions \cite{Brezin:1977gk,Stone:1977au,Bogomolny:1980ur,ZinnJustin:1981dx,Balitsky:1985in,Aoyama:1998nt,Unsal:2012zj,Aniceto:2013fka}.}
 \end{itemize}
 
Resurgent trans-series analysis resolves these two problems, producing an expression for the observable (such as energy) that is real and unambiguous.  This approach unifies the perturbative (P) series with a sum over all non-perturbative (NP)  contributions, forming a so-called ``trans-series'' expression, and the various terms in this trans-series are connected by an infinite ladder of intricate inter-relations which encode the cancellation of all imaginary and ambiguous terms \cite{Ecalle:1981,delabaere,Costin:2009}. We refer to this generalized notion of summability as {\it Borel-\'Ecalle summability.} 
For example, the leading ambiguous imaginary term arising from a  Borel analysis of the divergent perturbative series is of order $\pm i \pi e^{-2S_I/g^2}$.  This  is exactly cancelled by an identical term  
in the instanton-anti-instanton amplitude, $[{\cal I} \bar {\cal I}]_{\pm} \sim  e^{-2S_I/g^2}   \pm i \pi e^{-2S_I/g^2}$, whose  imaginary part is   also ambiguous, and  which lives in the non-perturbative part of the trans-series.  We refer to this cancellation mechanism 
in  quantum mechanics  as  Bogomolny-Zinn-Justin (BZJ) mechanism \cite{Bogomolny:1980ur,ZinnJustin:1981dx}.  A very  important aspect of  the theory of  ``resurgence'' is the statement that these cancellations  occur  to all NP-orders, including P-fluctuations around NP-saddles.  Thus the full trans-series is real, unique and unambiguous \cite{Aniceto:2013fka}.

This beautiful BZJ mechanism of cancellation of ambiguities  between non-Borel-summable perturbation theory and the non-perturbative multi-instanton sector has been explored in some detail for quantum mechanics (QM) problems with degenerate minima \cite{Brezin:1977gk,Stone:1977au,Bogomolny:1980ur,ZinnJustin:1981dx,Balitsky:1985in,h2plus,ddp,Aoyama:1998nt,zjj,Jentschura:2004jg}, but in fact this resurgent structure is a general property of perturbation theory that is also  relevant for quantum field theory (QFT), in particular  when there are degenerate classical vacua.  For example, in asymptotically free quantum field theories such as 4D $SU(N)$ gauge theory or 2D ${\mathbb C}{\mathbb P}^{N-1}$ theories there are infrared renormalons that lead to non-Borel summability of perturbation theory. This is a serious problem, because it means that perturbation theory on its own is ill-defined, just as is the case for the QM problems with degenerate minima. Until  recently it was not known how to cancel the resulting ambiguous imaginary parts against non-perturbative amplitudes, because for both 4D $SU(N)$  gauge theory and 2D ${\mathbb C}{\mathbb P}^{N-1}$, the IR renormalons lead to non-perturbative effects with exponential factors having exponents depending parametrically on $N$ as $2 S_{I}/N$, and such non-perturbative factors do not appear in these theories defined on ${\mathbb R}^4$ or ${\mathbb R}^2$, respectively \cite{'tHooft:1977am,David:1980gi}. However, a resolution of this problem has recently been proposed \cite{Argyres:2012ka,Dunne:2012ae}, motivated by another problem in the non-perturbative sector, which is that the instanton gas analysis (which works well for QM) is inconsistent for these QFT's defined on ${\mathbb R}^4$ or ${\mathbb R}^2$.   The dilute instanton gas approximation assumes that the inter-instanton separation is much larger than the size of the instanton, while  classical scale invariance implies that instantons of arbitrary  size come with  the same action  (leading to uncontrolled infrared divergences), 
hence the assumption is invalid.  A regularization of the QFT by spatial compactification (either twisting the boundary conditions or center-stabilizing deformation)  at weak coupling semi-classical regime
produce fractionalized instantons, ``molecules'' of which are associated with non-perturbative factors of the form $e^{ -2 S_{ I}/(g^2N) }$. This is 
appropriate for canceling the ambiguities from the semi-classical realization of IR renormalon singularities. For ${\mathbb C}{\mathbb P}^{N-1}$ models, the $N$ dependence matches precisely the $N$ dependence coming from the IR renormalons  \cite{Dunne:2012ae}, while for 4D gauge theory the dependence is parametrically correct\cite{Argyres:2012ka}.

Since this is a new type of QFT argument, using resurgent analysis to relate the IR renormalon problem of perturbation theory in asymptotically free theories with the IR divergence of the non-perturbative instanton gas, and trans-series expansions are still somewhat unfamiliar in much of the physics community, this paper is designed to be a simple pedagogical introduction to the physical origin of trans-series expansions. Our presentation is mainly in terms of two important quantum mechanical examples, the double-well and Sine-Gordon potentials, since these contain already much of the physics relevant for the discussion of non-perturbative effects due to degenerate minima in gauge theories and ${\mathbb C}{\mathbb P}^{N-1}$ models. In fact, these field theories can be continuously connected to the  quantum mechanical systems with periodic potentials.  
However, beyond our pedagogical presentation, we also make a new observation. For these theories (and others listed below), we show in explicit detail that: 
\begin{itemize}
\item { The perturbative series contains {\bf all} information about the non-perturbative sector, to all non-perturbative orders.}
\item{ Perturbation theory around the perturbative vacuum and fluctuations about all non-perturbative saddles (multi-instantons) are interrelated in a precise manner: high orders of fluctuations about one saddle are determined by low orders about "nearby" saddles (in the sense of action). }
\end{itemize}
 These are  extremely nontrivial facts, providing clear and direct illustrations of the surprising power of resurgent analysis. 
The first point was observed previously in the double-well system \cite{alvarez}, but here we show that the result is more general \cite{Dunne:2013ada}.

There is some  body of work concerning trans-series expansions for wave-functions, special functions and solutions to Schr\"odinger-like equations, as well as nonlinear differential equations \cite{Dingle:1973,BerryHowls,delabaere,Costin:2009}. Since we are motivated by attempts to compute QFT quantities such as a mass gap, to be very concrete we focus  on energy eigenvalues, rather than on wave-functions, but these approaches are obviously closely related. There is also an important set of ideas concerning exact quantization conditions \cite{dunham,voros,carlbook}, although these have mostly been investigated for QM potentials without degenerate vacua. We also stress that the basic idea of resurgent trans-series analysis is much more general, applying to both linear and nonlinear problems, and therefore should be applicable to functional problems like QFT, matrix models and string theory \cite{marino,Marino:2012zq,schiappa,Dunne:2012ae,Argyres:2012ka}.

\subsection{Where do the  trans-series  come from?}

In this paper we concentrate on trans-series expressions for energy eigenvalues in certain QM problems, with a coupling constant $g^2$. Our notation is chosen to match the coupling parameter $g^2$ in certain QFTs such as Yang-Mills or  ${\mathbb C}{\mathbb P}^{N-1}$ models. 
The general perturbative expansion of an energy level has the form
\begin{eqnarray}
E^{(N)}_{\rm pert. \, theory}(g^2)=\sum_{k=0}^\infty g^{2k} E_{k}^{(N)}
\label{pert1}
\end{eqnarray}
where $N$ is an integer labeling the energy level, and  the perturbative coefficients $E_{k}^{(N)}$ can be computed by straightforward iterative procedures. For the cases we study here, potentials with degenerate harmonic vacua, this perturbative expansion is not Borel summable, which means that {\it on its own} it is incomplete, and indeed inconsistent.

\begin{figure}[htb]
\includegraphics[angle=0, width=.99\textwidth]{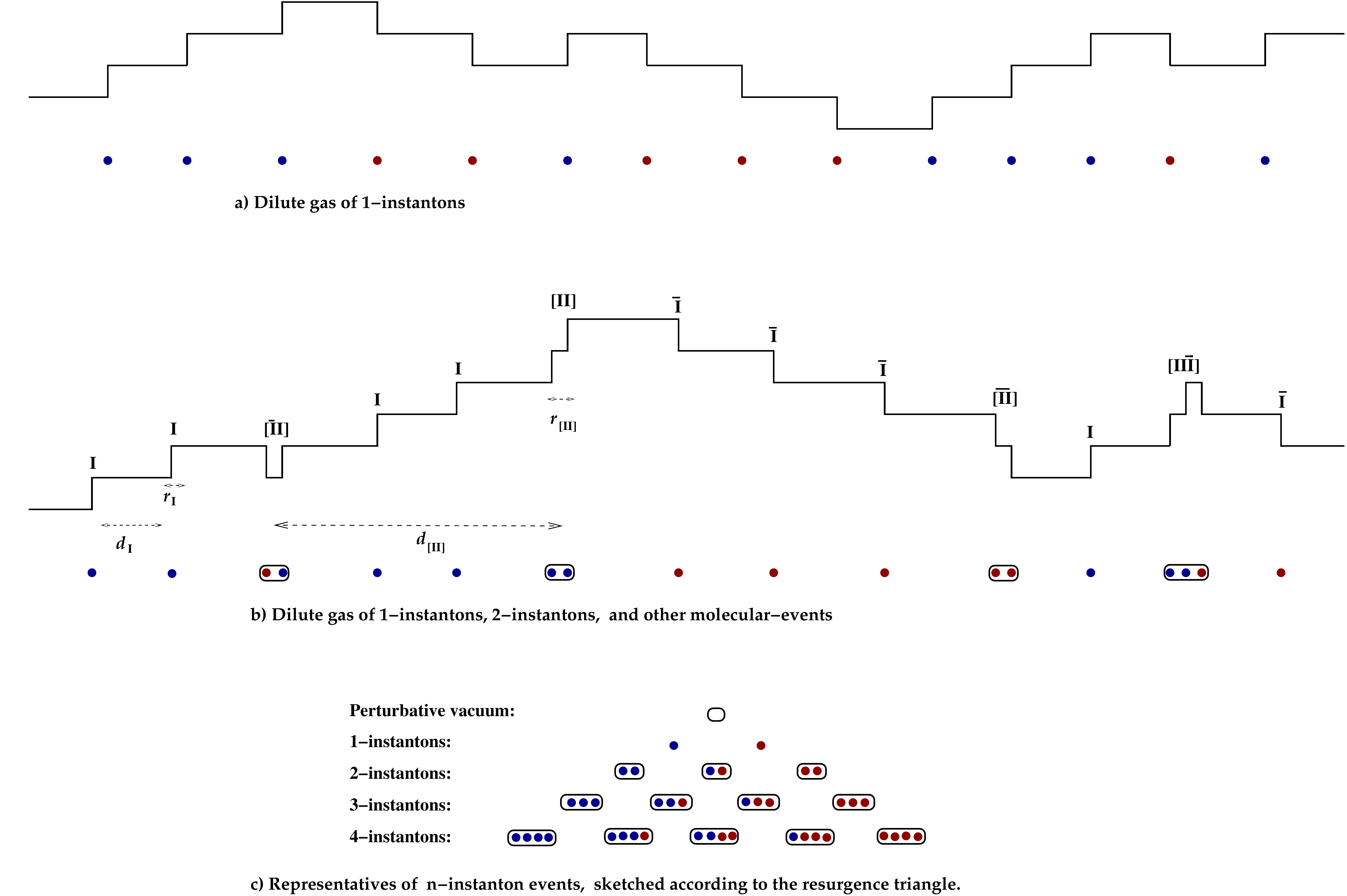}
\caption{   \textsf  {a) Dilute gas of  1-instantons  for a periodic potential (as given in typical textbook treatment). 
b) Dilute gas of 1-instantons, 2-instantons,  3-instantons, etc.   
2-instanton events (topological molecules) such as $[\cI \cI],  [\bar \cI \bar \cI], [\cI \bar \cI]$ are rarer, but  present. 
The amplitude associated with neutral 2-instantons or any other $k$-instanton with neutral 2-instanton subcomponent  are multi-fold ambiguous. 
This ambiguity cures 
the ambiguity of perturbation theory around the perturbative vacuum. 
c) $n$-instanton events classified according to  homotopy  (columns) and resurgence (refined structure in each column.) 
This picture is the result of uniform WKB and multi-instanton approach. }
}
\label{molecules-plots}
\end{figure}

This situation can be remedied by recognizing that the full expansion of the energy at small coupling is in fact of the ``trans-series'' form \cite{Ecalle:1981,h2plus,delabaere,Costin:2009,zjj,ddp,zinnbook}:
\begin{eqnarray}
E^{(N)}(g^2)=E^{(N)}_{\rm pert. \, theory}(g^2)+ \sum_{\pm} \sum_{k=1}^\infty \sum_{l=1}^{k-1}\sum_{p=0}^\infty 
\underbrace{ \left(\frac{1}{g^{2N+1}}\,\exp\left[-\frac{c} {g^2}\right]\right)^k }_{ \rm k-instanton}
\underbrace{ \left(\ln \left[\pm \frac{1}{g^2}\right]\right)^l}_{\rm quasi-zero mode}  \underbrace{c_{k, l, p}^{\pm}  g^{2p}}_{\rm perturbative \; fluctuations}
\label{trans}
\end{eqnarray}
In (\ref{trans}) we have artificially separated the perturbative expansion in the zero-instanton sector.  The second part of the trans-series involves a sum over powers of non-perturbative factors $\exp[- c/g^2]$, multiplied by prefactors that are themselves series in $g^2$ and in $\ln(\pm 1/g^2)$. The basic building blocks of the trans-series,  $g^2$, $\exp[-c/g^2]$  and $\ln(-1/g^2)$, are called ``trans-monomials'', and are  familiar from QM and QFT.  In physical terms, the trans-series is a sum over instanton contributions, with the  perturbative fluctuations about each instanton, and logarithmic terms coming from quasi-zero-modes.   A transseries  therefore 
combines perturbation theory with a dilute gas of 1-instantons, 2-instantons,  3-instantons, etc.\footnote{$n$-instanton is a  correlated $n$-event,  and  should be  distinguished from uncorrelated  $n$ 1-instanton events. }
    Note that in a typical textbook treatment, only the proliferation of 1-instanton events is accounted for. However, in order to make sense of ({\it i.e.} to define consistently) the  semi-classical expansion,  one needs to take into account a dilute gas of both 1-instanton as well as $k$-instantons, where $k \geq 2$.   See Fig.\ref{molecules-plots}
for a snap-shot of the Euclidean vacuum of the theory for the case of periodic potential. The sub-figure shows examples of 
$k$-instantons (molecular events).   

At second order in the instanton expansion,  quasi-zero-mode logarithms are first generated. 
Remarkably, the expansion coefficients $c_{k, l, p}$ of the trans-series are inter-twined amongst themselves, and also with the coefficients of the perturbative expansion, in such a way that the total trans-series is real and unambiguous. This intertwining can be represented graphically by the ``resurgence triangle'' introduced in \cite{Dunne:2012ae}, shown in Fig. \ref{molecules-plots}(c), and discussed in detail below for both the double-well and Sine-Gordon potentials.
For example, a Borel analysis of the perturbative series requires an analytic continuation in $g^2$, producing non-perturbative imaginary parts, but these are precisely cancelled by the imaginary parts associated with the $\ln(-1/g^2)$ factors in the non-perturbative portion of the trans-series. This applies not just at leading order, but to all subsequent orders arising from Borel summation of the divergent fluctuation expansion around any instanton sector. Ambiguities only arise if one  looks at just one isolated portion of the trans-series expansion, for example just the perturbative part, or just some particular multi-instanton sector. When viewed as a whole, the trans-series expression is unique and exact. 
We call this generalized summability of a non-Borel summable series Borel- \'Ecalle summability \cite{Ecalle:1981}. 

We have three main goals in this paper:  
\begin{enumerate}
\item
Explain in a simple manner how such a trans-series expansion (\ref{trans})  arises, and also in what sense  it is generic.
\item
Explain the origin of the  inter-relations within the trans-series, and their physical consequences.
\item
In its strongest form,  ``resurgence'' claims that complete knowledge of the perturbative series is sufficient to generate the remainder of the trans-series, including all orders of the non-perturbative expansion. We show here in simple and explicit detail how this works in QM systems with degenerate harmonic vacua.

\end{enumerate}

We comment that there is not yet universal agreement in the mathematical literature concerning the rigorous proof of the generality of trans-series expansions for resurgent functions. References \cite{ddp} contain proofs, but in a recent talk Kontsevich has raised questions about the rigor of  mathematical  results concerning resurgent functions \cite{kontsevich}. Nevertheless,  each of the trans-series monomials has a clear physical meaning, and here we show using relatively elementary techniques [uniform WKB] that the energy eigenvalues have precisely this
  trans-series structure in QM systems with degenerate harmonic vacua. Morover, these trans-monomial elements also have clear physical meaning in quantum field theory.

\section{Uniform WKB  For Potentials with Degenerate Minima}
\label{sec:uniform}

\subsection{The Spectral Problem}
Consider the spectral problem
\begin{eqnarray}
 -\frac{d^2}{dx^2}\psi(x)+V(x) \psi(x)=E\, \psi(x)
\label{spectral}
\end{eqnarray}
We are interested in cases where the potential $V(x)$ has degenerate minima, which are locally harmonic: $V(x)\approx x^2+ \dots$.
The two paradigmatic cases we study in detail are the double-well (DW) and Sine-Gordon (SG) potentials:
\begin{eqnarray}
V_{\rm DW}(x)&=&x^2(1+g\, x)^2=x^2+ 2g\, x^3+g^2 x^4
\label{vdw}\\
V_{\rm SG}(x)&=&\frac{1}{g^2}\sin^2(g\, x)=x^2-\frac{1}{3}g^2\, x^4+\dots
\label{vsg}
\end{eqnarray}
The Sine-Gordon case can be directly related to the Mathieu equation by simple changes of variables.
This permits detailed comparison with known results for Mathieu functions \cite{nist,goldstein}.

It is convenient to rescale the coordinate variable to $y=g\, x$:
\begin{eqnarray}
-g^4\, \frac{d^2}{dy^2}\psi(y)+V(y) \psi(y)=g^2\, E\, \psi(y)
\label{hbar}
\end{eqnarray}
where
\begin{eqnarray}
V_{\rm DW}(y)&=&y^2(1+y)^2
\label{vdw2}\\
V_{\rm SG}(y)&=&\sin^2(y)
\label{vsg2}
\end{eqnarray}
It is well known that in both these cases the perturbative energy levels are split by non-perturbative instanton effects. This level splitting is (at leading order) a single-instanton effect, and is textbook material \cite{landau,coleman,zinnbook,kenbook}.  From (\ref{hbar}) we see that $g^4$ plays the role of $\hbar^2$, and so we expect these non-perturbative effects to be characterized by exponential factors of the form
\begin{eqnarray}
\exp\left(-\frac{c}{g^2}\right)
\label{np}
\end{eqnarray}
for some constant $c>0$.

More interestingly, the perturbative series for these spectral problems is non-Borel-summable, and in the Borel-\'Ecalle approach is defined by the analytic continuation $g^2\to g^2\pm i\epsilon$, which induces a non-perturbative imaginary part, even though both potentials are completely stable and the energy should be purely real. As mentioned in the Introduction, the
resolution of this puzzle is  the Bogomolny-Zinn-Justin  (BZJ) mechanism: The non-perturbative imaginary part is in fact at the  two-instanton  order, and is canceled by a corresponding non-perturbative imaginary contribution coming from the instanton/anti-instanton amplitude \cite{Bogomolny:1980ur,ZinnJustin:1981dx,Balitsky:1985in,zinnbook}. 
The resurgent trans-series expression (\ref{trans}) for the energy eigenvalue encodes the fact that there is  an infinite tower of such cancelations, thereby relating properties of the perturbative sector and the non-perturbative sector. The BZJ cancelation is the first of this tower. A new observation we make here (see Section V)  is that we do not need to compute separately the perturbative and non-perturbative sectors: in fact, the perturbative series encodes {\bf all} information about the non-perturbative sector, to all non-perturbative orders.

\subsection{Strategy of the Uniform WKB Approach 
}
Before getting into details, we first state our strategy, and the basic result, which explains already why the expression for the energy eigenvalues has the trans-series form in (\ref{trans}). Since the potentials we consider have degenerate harmonic vacua,  in the $g^2\to 0$ limit each classical vacuum has the form of a harmonic oscillator well. Therefore it is natural to use a parabolic {\it uniform WKB} ansatz for the wave-function \cite{langer,cherry,alvarez,millergood,galindo}: 
\begin{eqnarray}
\psi(y) = \frac{D_\nu \left(\frac{1}{g}u(y)\right)}{\sqrt{u^\prime(y)}}
\label{uniform}
\end{eqnarray}
Here  $D_\nu$ is a parabolic cylinder function \cite{nist} (the solution to the harmonic problem), and $\nu$ is  an ansatz parameter that is to be determined. When $g^2=0$ we would have an isolated harmonic well, and $\nu$ would be an integer $N$. For $g^2>0$, we find that $\nu$ is close to an integer [see (\ref{global}) below].

Substituting this uniform WKB ansatz form (\ref{uniform}) of the wave-function into the 
Schr\"odinger equation (\ref{hbar}) produces a nonlinear equation for the argument function $u(y)$, and this equation can be solved perturbatively. Purely local analysis in the immediate vicinity of the potential minimum, where the potential is harmonic, leads to a perturbative expansion of the energy (explained in Section \ref{pert-uniform} below):
\begin{eqnarray}
E=E(\nu, g^2)=\sum_{k=0}^\infty g^{2k} E_k(\nu)
\label{pert}
\end{eqnarray}
The coefficient $E_k(\nu)$ depends on the as-yet-undetermined ansatz parameter $\nu$. In fact, $E_k(\nu)$ is a polynomial in $\nu$, of degree $(k+1)$. In the $g^2\to 0$ limit, the ansatz parameter $\nu$ tends to an integer $N$,  labelling the unperturbed harmonic oscillator energy level. Indeed, when $\nu=N$, the expansion (\ref{pert}) coincides precisely with standard Rayleigh-Schr\"odinger perturbation theory:
\begin{eqnarray}
E\left(\nu=N, g^2\right)\ \equiv  E_{\rm pert.\, theory}^{(N)}(g^2)
\label{equiv1}
\end{eqnarray}
This perturbative series expression is incomplete, and indeed ill-defined, because the series is not Borel summable. The fact that it is incomplete should not be too surprising because so far the analysis has been purely {\it local}, making no reference to the existence of neighboring degenerate classical vacua. To fully determine the energy we must impose a {\it global} boundary condition that relates one classical vacuum to another. When we do this we learn that $\nu$ is only exponentially close to the integer $N$, with a small correction $\delta\nu$ that is a function of both $N$ and $g^2$:
\begin{eqnarray}
\nu_{\rm global}(N, g^2)=N+\delta\nu(N, g^2)
\label{global}
\end{eqnarray}
The explicit form of the correction term $\delta\nu(N, g^2)$ is derived and discussed below in Section \ref{sec:global}. For now we  state that generically it has a trans-series form:
\begin{eqnarray}
\delta\nu(N, g^2)=\sum_\pm\sum_{k=1}^\infty \sum_{l=1}^{k-1}\sum_{p=0}^\infty d^{(\pm)}_{k, l, p} 
\left(\frac{1}{g^{2N+1}}\,\exp\left[-\frac{c}{g^2}\right]\right)^k 
\left(\ln \left[\mp\frac{1}{g^2}\right]\right)^l g^{2p}
\label{f}
\end{eqnarray}
We show in Section \ref{sec:global} that this form  follows directly from properties of the parabolic cylinder functions, and so it is generic  to problems having degenerate vacua that are harmonic.
\footnote{For a curious counter-example to the oft-held belief that non-Borel-summable expansions occur for any potential with degenerate vacua, consider the {\it non-harmonic} case of two square wells, separated by a distance $1/g$, and with central barrier of height $1/g^2$. This $g$ dependence is chosen to mimic that of the double-well potential. This is an elementary  problem, soluble in terms of hyperbolic trigonometric functions, and an expansion of the eigenvalue condition for small $g$ produces a trans-series expansion, but without any $\ln(-1/g^2)$ terms. Moreover,  one finds that the ``perturbative'' small $g^2$ expansion is in fact summable. Thus, the trans-series structure is quite different in this non-harmonic case. One could argue that this case is ill-defined because the bottom of each well is flat, so there is no real vacuum location, but the same conclusion can be obtained by replacing the square wells by triangular wells, which is also a soluble problem, in terms of Airy functions. Periodic versions of these cases  also produce interesting trans-series. Thus, the harmonic nature of the classical vacua is a significant feature of the argument.} 

Having solved the global boundary condition to determine the parameter $\nu$ as a function of $N$ and $g^2$, as in (\ref{global}) and (\ref{f}), to obtain the corresponding energy eigenvalue we insert this value $\nu_{\rm global}(N, g^2)$ back into the perturbative expansion (\ref{pert}) for the energy, leading to the final exact expression for the energy eigenvalue:
\begin{eqnarray}
E^{(N)}(g^2)=E\left(N+\delta\nu(N, g^2), g^2\right)=\sum_{k=0}^\infty g^{2k} E_k(N+\delta\nu(N, g^2))
\label{answer}
\end{eqnarray}
Re-expanding the polynomial coefficients $E_k(N+\delta\nu(N, g^2))$ for small coupling $g^2$, we obtain the trans-series  expression (\ref{trans}) for the $N^{\rm th}$ energy level, $E^{(N)}(g^2)$. 
\begin{eqnarray}
E_N(g)=E(N, g)+(\delta \nu) \left[\frac{\partial E}{\partial \nu}\right]_{N}
+\frac{(\delta \nu)^2}{2} \left[\frac{\partial^2 E}{\partial \nu^2}\right]_{N}+\dots
\label{answer-2}
\end{eqnarray}

We stress that this uniform WKB approach makes it very clear why the trans-series form of the energy is generic for problems with degenerate harmonic classical vacua: all properties of the $g^2\to 0$ limit reduce to properties of the parabolic cylinder functions, which lead directly to the trans-series form for $\delta\nu(N, g^2)$ in (\ref{f}). In particular, all analytic continuations needed to analyze questions of resurgence and cancellation of ambiguities can be expressed in terms of the known analytic continuation properties of the parabolic cylinder functions \cite{nist}.

\subsection{Perturbative Expansion of the Uniform WKB Ansatz}
\label{pert-uniform}

Recalling that the parabolic cylinder function $D_\nu(z)$ satisfies the differential equation \cite{nist}
\begin{eqnarray}
\frac{d^2}{dz^2}D_\nu(z)+\left(\nu+\frac{1}{2}-\frac{z^2}{4}\right)D_\nu(z)=0
\label{pcf}
\end{eqnarray}
we see that the uniform WKB ansatz (\ref{uniform}) converts the Schr\"odinger equation (\ref{hbar}) to the following non-linear equation for the argument function $u(y)$ appearing in (\ref{uniform}):
\begin{eqnarray}
V(y)-\frac{1}{4}u^2(u^\prime)^2-g^2\, E+g^2 \left(\nu+\frac{1}{2}\right)(u^\prime)^2+\frac{g^4}{2}\sqrt{u^\prime}\left(\frac{u^{\prime\prime}}{(u^\prime)^{3/2}}\right)^\prime=0
\label{nle}
\end{eqnarray}
Here $u^\prime$ means $du/dy$.
At first sight, it looks like (\ref{nle}) is more difficult to solve than the original Schr\"odinger equation (\ref{hbar}), but we will see that the perturbative solution of (\ref{nle}) has some advantages over the  perturbative solution of (\ref{hbar}). We solve (\ref{nle}) for $u(y)$ and $E$ by making simultaneous perturbative expansions:
\begin{eqnarray}
E&=& E_0+g^2 E_1+g^4 E_2+\dots
\label{pe1}\\
u(y)&=& u_0(y)+g^2\, u_1(y)+g^4\, u_2(y)+\dots
\label{pu1}
\end{eqnarray}
Note that the expansion coefficients $E_k$ and $u_k(y)$ also depend on the as-yet-undetermined parameter $\nu$ that appears in the ansatz (\ref{uniform}), and consequently in the equation (\ref{nle}). This parameter $\nu$ is not determined by the local perturbative expansions in (\ref{pe1}, \ref{pu1}); the parameter $\nu$ requires global non-perturbative information describing how one perturbative vacuum potential well  connects to another. This is discussed below in Section \ref{sec:global}.

\subsubsection{Leading Order: 
 Origin of the Usual Exponential WKB  Factor}
At zeroth order in $g^2$ the equation (\ref{nle}) implies:
\begin{eqnarray}
u_0^2(u_0^\prime)^2=4V\qquad  \Rightarrow \qquad 
u_0^2(y)=4\int_0^y dy\,\sqrt{V}
\label{zeroth}
\end{eqnarray}
where the lower limit is chosen to satisfy the small $y$ limiting behavior of the non-linear equation (\ref{nle}). In particular, since each well is locally harmonic, $V(y)\approx y^2$, we learn that 
\begin{eqnarray}
u_0(y)\approx \sqrt{2}\, y +\dots  \qquad, \qquad y\to 0
\label{limit1}
\end{eqnarray}
Correspondingly, the $O(g^2)$ term in (\ref{nle}) then tells us that the perturbative expansion for the energy begins as
\begin{eqnarray}
E=(2\nu+1)+\dots
\label{limit2}
\end{eqnarray}
The results (\ref{limit1}, \ref{limit2}) are simply  reflections of the locally harmonic nature of the $g^2\to 0$ limit.

For  the DW and SG potentials, (\ref{zeroth}) yields:
\begin{eqnarray}
{\rm DW}: u_0(y)= \sqrt{2}\, y\,  \sqrt{1+\frac{2 y}{3}}\qquad; \qquad
{\rm SG}: u_0(y)=2\sqrt{2}\, \sin\left(\frac{y}{2}\right)
\label{u0}
\end{eqnarray}
From the asymptotic behavior of the parabolic cylinder function \cite{nist}, in the $g^2\to 0$ limit we find the expected exponential WKB  factor:
\begin{eqnarray}
D_\nu(z)\sim z^\nu \, e^{-z^2/4}\quad, \quad (z\to +\infty) \quad \Rightarrow \quad \psi(y)\sim \exp\left[- \frac{u_0^2}{4g^2}\right]\sim \exp\left[- \frac{1}{g^2}\int_0^y \sqrt{V}\right]
\label{wkb1}
\end{eqnarray}
We discuss the pre-factors below in Section \ref{sec:resurgence}.  \footnote{
The relation between the uniform WKB wave-function and instanton amplitude is following:
In our normalization of Hamiltonian (\ref{spectral}), $m=\frac{1}{2}$.  Thus, 
 the BPS-bound for the instanton action is  
$ S[y]=\frac{1}{g^2}  \int dt \left( \frac{1}{4} \dot y^2 + V(y) \right) \geq 
\frac{1}{g^2}  \int_{0}^{ 2y_{\rm mid-point} } dy \sqrt V =\frac{S_I}{g^2}$
  Thus, according to (\ref{wkb1}), the leading 
uniform WKB wave-function at $y_{\rm mid-point}$ is therefore 
$ \psi\left(y_{\rm mid-point} \right) \sim e^{-S_I/(2 g^2)}$, and is exponentially small, i.e, the value of the uniform WKB wave function at the midpoint between the two harmonic  minima  (see Fig.~\ref{dw-plots} or Fig.~\ref{sg-plots}) is square root of instanton fugacity. Also see the discussion around  (\ref{xidw}).}

\subsubsection{Higher Orders } 

The higher-order perturbative  solution is straightforward but tedious. Imposing the boundary condition of finiteness of $u^2(y)$ at $y=0$, one finds that the energy $E(\nu, g^2)$ has an expansion of the form
\begin{eqnarray}
E(B, g^2)=2B-\sum_{k=1}^\infty g^{2k} p_{k+1}(B)\qquad, \qquad B\equiv \nu+\frac{1}{2}
\label{pte}
\end{eqnarray}
where it proves convenient to express the coefficients in terms of the parameter $B\equiv \nu+\frac{1}{2}$. The leading term is universal [recall (\ref{limit2})], and the coefficients, $p_{k+1}(B)$, of this expansion are polynomials of degree $(k+1)$ in $B$. Moreover, they have definite parity: $p_k(-B)=(-1)^k p_k(B)$. For example, in the two explicit cases of the double-well (DW) and Sine-Gordon (SG) potentials:
\begin{eqnarray}
E_{\rm DW}(B, g^2)&=&2B-2g^2\left(3B^2+\frac{1}{4}\right)-2g^4\left(17 B^3+\frac{19}{4}B\right)-2g^6\left(\frac{375}{2}B^4+\frac{459}{4}B^2+\frac{131}{32}\right)\nonumber\\
&&
-2g^8\left(\frac{10689}{4}B^5+\frac{23405}{8}B^3+\frac{22709}{64}B\right)-\dots
\label{ebdw0}
\\
E_{\rm SG}(B, g^2)&=&2B-\frac{g^2}{2}\left(B^2+\frac{1}{4}\right)
-\frac{g^4}{8}\left(B^3+\frac{3}{4}B\right) 
-\frac{g^6}{32}\left(\frac{5}{2}B^4+\frac{17}{4}B^2+\frac{9}{32}\right) \nonumber\\
   &&-\frac{g^8}{128}\left(\frac{33}{4}B^5+\frac{205}{8}B^3+\frac{405}{64}B\right) -\dots
\label{ebsg}
\end{eqnarray}
An important observation is that if we replace $\nu$ by an integer quantum number $N$, so that $B= N+\frac{1}{2}$, then the expansions (\ref{pte}, \ref{ebdw0}, \ref{ebsg}) coincide precisely with the corresponding expansion obtained from standard Rayleigh-Schr\"odinger perturbation theory about the $N^{\rm th}$  harmonic oscillator level:
\begin{eqnarray}
E\left(B=N+\frac{1}{2}, g^2\right)\ = E_{\rm pert.\, theory}^{(N)}(g^2)
\label{equiv}
\end{eqnarray}
In particular, note that for each $B>0$, the perturbative expansion in $g^2$, as in (\ref{pte}, \ref{ebdw0}, \ref{ebsg}), is a divergent non-alternating series, which is not Borel summable. This fact will be crucial below when we come to discuss the global boundary conditions that connect one perturbative vacuum to another: see Section \ref{sec:global}.

The corresponding perturbative expansion for $u(y)$ [the function that appears in the argument of the parabolic cylinder function in the uniform WKB ansatz (\ref{uniform})] is of the form:
\begin{eqnarray}
u(y)=u(y, B, g^2)=\sum_{k=0}^\infty g^{2k} \, u_k(y, B)
\label{ptu}
\end{eqnarray}
With respect to its dependence on  $B$, 
the coefficient function $u_{k}(y, B)$ is a polynomial of degree $k$ in $B$, with definite parity: $u_k(y, -B)=(-1)^k u_k(y, B)$. For  the DW and SG cases:
\begin{eqnarray}
u_{\rm DW}(y)&=& \sqrt{2}\, y\,  \sqrt{1+\frac{2 y}{3}}
+g^2\, B\, \frac{\ln\left[ \left(1+\frac{2 y}{3}\right)(1+y)^2\right]}{ \sqrt{2}\, y \sqrt{1+\frac{2 y}{3}}}
+\dots 
\label{updw}\\
u_{\rm SG}(y)&=&2\sqrt{2}\sin \frac{y}{2}+g^2\, B\,  \frac{\ln \cos \frac{y}{2}}{\sqrt{2}\, \sin \frac{y}{2}}+\dots 
\label{upsg}
\end{eqnarray}
Higher order terms are straightforward to generate but cumbersome to write. 

While the perturbative expansion (\ref{pe1}) of the energy yields, with the identification $\nu\to N$, exactly the same perturbative series for the energy eigenvalue as Rayleigh-Schr\"odinger perturbation theory [see (\ref{equiv})],  the situation is quite different for  the wave-function expansion in (\ref{pu1}, \ref{ptu}). To recover the Rayleigh-Schr\"odinger perturbation theory wave-function for the $N^{\rm th}$ level, we use the uniform WKB ansatz (\ref{uniform}), identify $\nu\to N$, rewrite $y=g\, x$, and expand in $g^2$:
\begin{eqnarray}
\psi^{(N)}(x) &=& \frac{D_N \left(\frac{1}{g}\left[u_0(g\, x)+g^2 u_1(g\, x)+g^4 u_2(g\, x)+\dots\right]\right)}{\sqrt{(d/dx)\left[ u_0(g\, x)+g^2 u_1(g\, x)+ g^4 u_2(g\, x)+\dots\right]/g}} \nonumber\\
&
\equiv& \frac{D_N(\sqrt{2} x)}{\sqrt{2}}+g^2 \psi_1^{(N)}(x)+g^4 \psi_2^{(N)}(x)+\dots
\label{wf}
\end{eqnarray}
The leading term is the familiar harmonic oscillator wave function for the unperturbed $N^{\rm th}$ level. Interestingly, if we truncate the perturbative expansion of $u(g\, x)$ at some order $g^{2k}$, and use this {\it inside} the uniform WKB expression (\ref{uniform}), we obtain a much better approximation to the wave-function than the  truncation of the Rayleigh-Schr\"odinger perturbation theory wave-function at the same order $g^{2k}$. The uniform WKB approximation effectively gives a resummation of many orders of Rayleigh-Schr\"odinger perturbation theory.

\section{Global boundary conditions}
\label{sec:global}

\subsection{Relating one minima to another}

So far the entire discussion has been local, in the neighborhood of the minimum of one of the classical vacua. To proceed, we need to specify how one classical vacuum relates to another. Here the details of the double-well and Sine-Gordon cases differ slightly, but in each case we impose a global boundary condition at the midpoint of the barrier between two neighboring classical vacua (we restrict ourselves here to symmetric barriers). The result illustrates the physics of level splitting (DW) and band spectra (SG), respectively.

\begin{figure}[htb]
\includegraphics[scale=1]{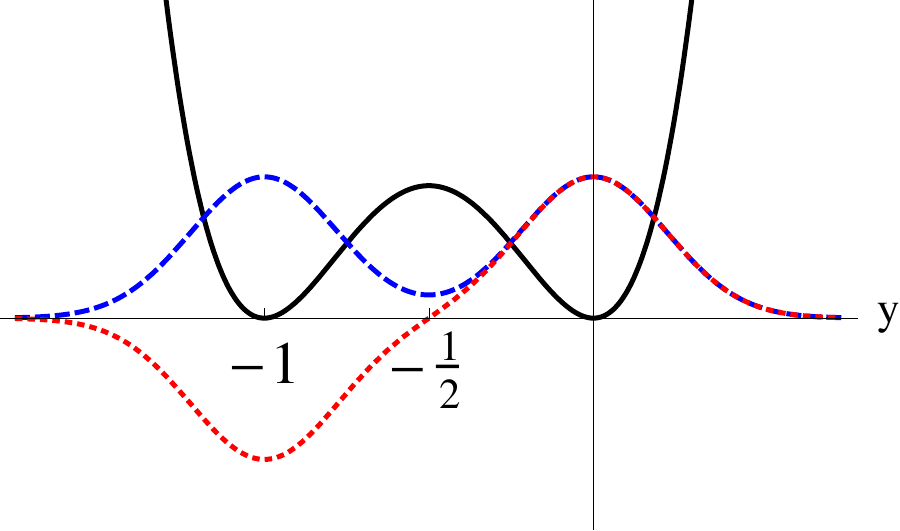}
\caption{ \textsf{ The global boundary condition for the lowest two states in the double-well potential $V(y)=y^2(1+y)^2$. The lower state wave function is nodeless and has vanishing derivative at the midpoint of the barrier. The upper state wave function has one node at the midpoint of the barrier.}}
\label{dw-plots}
\end{figure}

Consider first the DW potential. Each  level labeled by the index $N$ splits into two levels due to tunneling between the two classical vacua. To see how this arises, consider $N=0$ and note that the ground state wave-function is a node-less function, which is therefore an even function about the midpoint between the two wells ($y_{\rm midpoint}=-\frac{1}{2}$), while the first excited state wave-function (which also has $N=0$) has one node and  is therefore an odd function about this midpoint: see Figure \ref{dw-plots}. Thus, the global boundary condition to be imposed at $y_{\rm midpoint}$ is:
\begin{eqnarray}
&&{\rm ground\, state:}\qquad \psi_{\rm DW}^{\,\prime}\left(-\frac{1}{2}\right)=0
\label{gdw1}\\
&&{\rm first\, excited\, state:}\qquad \psi_{\rm DW}\left(-\frac{1}{2}\right)=0
\label{gdw2}
\end{eqnarray}
Because of the reflection symmetry of the DW potential about the midpoint,  in effect we only need to solve the DW problem in the right-hand half-space, $-\frac{1}{2}\leq y < \infty$, with either a Neumann (ground state) or Dirichlet (first excited state) boundary condition at $y= -\frac{1}{2}$, and in both cases with a Dirichlet boundary condition at $y=+\infty$.
For higher energy levels (i.e., for higher values of $N$), we interchange the Neumann or Dirichlet boundary conditions at $y= -\frac{1}{2}$, according to whether $N$ is odd or even.

For the SG potential,  each perturbative level labeled by the index $N$ splits into a continuous band of states. This phenomenon arises from the Bloch condition: $\psi(y+\pi)=e^{i\,\theta}\psi(y)$, where $\theta$ is a real angular parameter $\theta\in [0, \pi]$ that labels states in a given band of the spectrum. This Bloch boundary condition is efficiently expressed in terms of the discriminant, using the standard Floquet analysis \cite{kenbook,nist}. Define two independent solutions $w_{I}(y)$ and $w_{II}(y)$, normalized as follows at some arbitrary chosen point (which we take here to be at $y=-\frac{\pi}{2}$, the center of a barrier between two classical vacua, as in the DW case):
\begin{eqnarray}
\begin{pmatrix}
w_I\left(-\frac{\pi}{2}\right) & w_I^\prime \left(-\frac{\pi}{2}\right) \cr
w_{II}\left(-\frac{\pi}{2}\right) & w_{II}^\prime \left(-\frac{\pi}{2}\right)
\end{pmatrix}
=
\begin{pmatrix}
1& 0\cr
0 & 1
\end{pmatrix}
\label{basis}
\end{eqnarray}
The Bloch condition is expressed in terms of the discriminant, which is itself expressed in terms of the functions $w_I$ and $w_{II}^\prime$ evaluated at a location shifted by one period, for example at $y=+\frac{\pi}{2}$:
\begin{eqnarray}
\cos\theta
&=&\frac{1}{2}\left(w_I\left(\frac{\pi}{2}\right) +w_{II}^\prime \left(\frac{\pi}{2}\right)\right)
\label{bloch1}\\
&=&
w_I\left(\frac{\pi}{2}\right)
\label{bloch}
\end{eqnarray}
\begin{figure}[htb]
\includegraphics[scale=1]{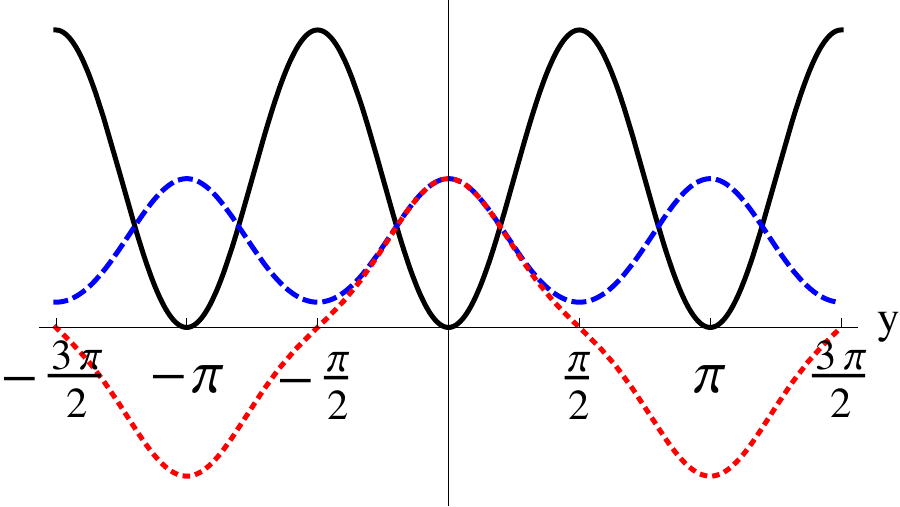}
\caption{\textsf{  The global boundary condition for the band-edge states of the lowest band for the Sine-Gordon potential $V(y)=\sin^2 y$. The lower band-edge wave function is nodeless and has vanishing derivative at the midpoint of each barrier. The upper band-edge wave function has one node at the midpoint of each barrier.}}
\label{sg-plots}
\end{figure}
In the last step we have used the symmetry of the SG potential which implies that  $w_{II}^\prime \left(\frac{\pi}{2}\right)=w_{I} \left(\frac{\pi}{2}\right)$, in order to write 
the Bloch condition in the compact form (\ref{bloch}). The band edge wave-functions are either periodic or anti-periodic functions of $y$, with period $\pi$, depending on whether $N$ is even or odd. For example, for the $N=0$ perturbative level, the wave-function  for the lower edge of the resulting band is a periodic function, while for the upper edge it is an anti-periodic function. See  Figure \ref{sg-plots}.

\subsection{Global Boundary Conditions in the Uniform WKB Approach}

Since the uniform WKB approximation (\ref{uniform}) to the wave function is expressed in terms of  parabolic cylinder functions, the implementation  of the global boundary condition in this approach  is intimately related to the properties of the parabolic cylinder functions. Moreover, since the argument of the parabolic cylinder function in the uniform ansatz (\ref{uniform}) goes like $u_0(y)/g$ in the $g^2\to 0$ limit, and $u_0(y_{\rm midpoint})$ is finite, we see that the global boundary condition in the $g^2\to 0$ limit is directly related to the asymptotic behavior of the parabolic cylinder functions at large values of their argument. It is at this stage that we must  confront the fact that the perturbative expansion of the energy in (\ref{pte}), and also the perturbative expansion of the function $u(y)$ in (\ref{ptu}), are in fact non-Borel-summable divergent series in $g^2$. This is because $g^2>0$ is a Stokes line, and we encounter the familiar problems of trying to make a perturbative expansion on a Stoke line \cite{Dingle:1973,BerryHowls,voros,budden,heading,carlbook}.  The theory of Borel-\'Ecalle resurgent summation provides a well-defined approach to this problem:
\begin{enumerate}
\item
Analytically continue in $g^2$ off the positive real axis. Then all the divergent series become Borel summable. This is often expressed \cite{Bogomolny:1980ur,ZinnJustin:1981dx,Balitsky:1985in,zinnbook} as continuing all the way to $g^2\to -g^2$, in which case the non-alternating non-Borel-summable series become alternating and Borel summable. In fact, it is enough to go slightly off the positive real $g^2$  axis: $g^2\to g^2\pm i\epsilon$, which avoids the Borel poles and/or branch cuts.
\item
Having obtained the Borel summed expressions, analytically continue in $g^2$ back to the positive real axis.
\item
This procedure produces non-perturbative imaginary contributions as the Borel summed series are continued back to the positive real $g^2$ axis; moreover, the overall sign of such a term is ambiguous, depending on whether one approaches the positive real $g^2$ axis from above or below. The remarkable fact is that if one makes  all analytic continuations consistently in the global boundary condition, then in the trans-series expansion all ambiguities in the perturbative expansions  are strictly correlated with corresponding ambiguities in the non-perturbative sectors, in such a way that all ambiguities cancel, producing an exact and unambiguous trans-series expression for the energy eigenvalue. 

\end{enumerate}

\subsection{Global Boundary Condition for the Double-Well System}

To derive the explicit form of the global boundary condition, recall that the global boundary conditions (\ref{gdw1}, \ref{gdw2}) are imposed at the barrier midpoint $y_{\rm midpoint}=-\frac{1}{2}$. When we analytically continue $g^2$ off the positive real axis, this renders   the $g^2$ expansion (\ref{ptu}) of the argument $\frac{1}{g}u\left(-\frac{1}{2}\right)$ of the parabolic cylinder function $D_\nu$ appearing in the uniform WKB ansatz (\ref{uniform}) Borel summable.  But now this argument  $\frac{1}{g}u\left(-\frac{1}{2}\right)$ is also a complex number, off the real positive axis. Thus in the limit where the modulus of $g^2$ approaches zero, the appropriate asymptotic behavior of the parabolic cylinder function is not just given by $D_\nu(z)\sim z^\nu e^{-z^2/4}\,,\, (z\to +\infty)$, as used in (\ref{wkb1}). We now need to use the (resurgent) asymptotic behavior of the parabolic cylinder functions throughout the relevant region of the complex plane, given by \cite{nist}:
\begin{eqnarray}
D_\nu(z)\sim z^\nu\, e^{-z^2/4}F_1\left(z^{2}\right)+e^{\pm i\pi\nu} \frac{\sqrt{2\pi}}{\Gamma(-\nu)} z^{-1-\nu}\, e^{z^2/4} F_2\left( z^{2}\right)\qquad , \qquad \frac{\pi}{2} < \pm \,{\rm arg}(z) <\pi
\label{full}
\end{eqnarray}
where 
\begin{eqnarray}
F_1\left(z^2 \right)&=&\sum_{k=0}^\infty \frac{\Gamma\left(k-\frac{\nu}{2}\right) \Gamma\left(k+\frac{1}{2}-\frac{\nu}{2}\right)}{\Gamma\left(-\frac{\nu}{2}\right) \Gamma\left(\frac{1}{2}-\frac{\nu}{2}\right)}\, \frac{1}{k!}\left(\frac{-2}{z^2}\right)^k \\
F_2\left( z^{2}\right)&=&\sum_{k=0}^\infty \frac{\Gamma\left(k+\frac{1}{2}+\frac{\nu}{2}\right) \Gamma\left(k+1+\frac{\nu}{2}\right)}{\Gamma\left(\frac{1}{2}+\frac{\nu}{2}\right) \Gamma\left(1+\frac{\nu}{2}\right)}\, \frac{1}{k!}\left(\frac{2}{z^2}\right)^k 
\label{fs}
\end{eqnarray}
Notice that there are two different exponential terms $e^{\pm z^2/4}$ in (\ref{full}). Normally  one or other is dominant or sub-dominant, but for certain rays of $z^2$ in the complex plane they may be equally important. This is a manifestation of the Stokes phenomenon \cite{Dingle:1973,BerryHowls,budden,heading,carlbook}.

Consider first the global boundary condition  with Dirichlet boundary condition at the midpoint, as in (\ref{gdw2}). Using the full analytic expression (\ref{full}), the global boundary condition (\ref{gdw2}) can be written as 
\begin{eqnarray}
&&\psi_{\rm DW}\left(-\frac{1}{2}\right)=0 \Longrightarrow D_\nu \left(\frac{u \left(-\frac{1}{2}\right)}{g}\right) =0  \qquad  {\rm for \; arbitrary}  \arg (g^2). \; \; {\rm Hence}  \cr
&& \frac{1}{\Gamma(-\nu)}\left(\frac{e^{\pm i\pi}\, 2}{g^2}\right)^{-\nu}=-\xi \, H_0(\nu, g^2)\quad; \quad (\text{upper level})
\label{dw2}
\end{eqnarray}
where the ``instanton factor'' is related to the zeroth order uniform WKB wave-function at the 
mid-point:
\begin{eqnarray}
\xi &\equiv& \sqrt{\frac{1}{\pi\, g^2}}\exp\left[-\frac{u_0^2\left(-\frac{1}{2}\right)}{2\, g^2}\right] = \frac{1}{\sqrt{\pi g^2}}\exp\left[-\frac{1}{6g^2}\right] =  \frac{1}{\sqrt{\pi g^2}}\exp\left[-\frac{S_I}{g^2}\right] 
\label{xidw}
\end{eqnarray}
and the perturbative ``fluctuations around the instanton'' are given by the function
\begin{eqnarray}
H_0(\nu, g^2)&\equiv &
  \left(\frac{u^2\left(-\frac{1}{2}\right)}{2}\right)^{\nu+\frac{1}{2}}
\frac{F_1\left(\frac{u^2\left(-\frac{1}{2}\right)}{g^2}\right)}{F_2\left(\frac{u^2\left(-\frac{1}{2}\right)}{g^2}\right)}\exp\left[-\frac{1}{2g^2}\left(u^2\left(-\frac{1}{2}\right)-u_0^2\left(-\frac{1}{2}\right)\right)\right]
\label{xih}
\end{eqnarray}
The form of (\ref{dw2}) follows directly from the global boundary condition (\ref{gdw2}) and the asymptotic properties (\ref{full}) of the parabolic cylinder functions.

The expression (\ref{dw2}) is an implicit relation for $\nu$ as a function of the coupling $g^2$. As $g^2\to 0$, it is clear that $\nu$ is close to a  non-negative  integer $N$. We solve by expanding  $\nu=N+\delta \nu$,  noting that
\begin{eqnarray}
\frac{1}{\Gamma(-\nu)}\left(\frac{e^{\pm i\pi}\, 2}{g^2}\right)^{-\nu}
=-N! \left(\frac{e^{\pm i\pi}\, 2}{g^2}\right)^{-N}\left\{ \delta \nu-\left[\gamma+\ln\left(\frac{e^{\pm i\pi}\, 2}{g^2}\right)-h_N\right](\delta\nu)^2+\dots\right\}
\label{dw3}
\end{eqnarray}
where $h_N$ is the $N^{\rm th}$ harmonic number \cite{nist} and $\gamma$ is  Euler-Mascheroni constant.  This implies that (for the odd state)
\begin{eqnarray}
\nu &=&N+  \frac{\left(\frac{2}{g^2}\right)^N H_0(N, g^2)}{N!}\,\xi 
-  \frac{\left(\frac{2}{g^2}\right)^{2N}}{(N!)^2}
\left[ H_0 \frac{\partial H_0}{\partial N} 
 +\left(\ln\left(\frac{e^{\pm i\pi}\, 2}{g^2}\right)-\psi(N+1)\right)H_0^2\right]  \xi^2 + O(\xi^3) \qquad 
\label{nu-exp}
\end{eqnarray}
This expansion is the trans-series form of the parameter $\nu$ mentioned already in (\ref{global})-(\ref{f}) in Section \ref{sec:uniform}. This discussion makes it clear that the ``instanton'' exponential factor $\xi$, the logarithmic factors, and the powers of $g^2$ all come from the expansion of the gamma function and the exponential factor in (\ref{dw2}), which ultimately  originate in the asymptotic form of the parabolic cylinder function (\ref{full}). This explains why the trans-series form  for the energy eigenvalue is generic for problems with degenerate harmonic vacua.

Notice that the leading imaginary part in (\ref{nu-exp}) occurs at $O(\xi^2)$, showing that it is generically a two-instanton effect, and moreover it is directly related to the square of the real part at $O(\xi)$:
\begin{eqnarray}
{\rm Im} \left[\nu-N\right]=\pm \pi\left( {\rm Re}\left[\nu-N\right]\right)^2 +  O(\xi^3)
\label{disp}
\end{eqnarray}
This is the first of a set of {\it confluence equations} \cite{Dunne:2012ae}, as discussed below in Section \ref{sec:resurgence}. The $\pm$ sign here comes from the ambiguity in the analytic continuation of $g^2$; it will be shown to be correlated with the ambiguity in the Borel summation of the perturbative series, in such a way that the ambiguous imaginary parts cancel.

If we repeat this argument using the Neumann boundary condition at the midpoint, then after some computation we find that the only change is a change in sign on the RHS of (\ref{dw2}), which leads to a change of sign of the odd powers of $\xi$ in (\ref{nu-exp}). Thus, to leading order in the exponentially small instanton factor $\xi$,  using (\ref{answer-2}) and (\ref{nu-exp}),
 the splitting of the levels is symmetric \footnote{At higher orders in $\xi$ the  splitting is no longer symmetric.}:
\begin{eqnarray}
E^{(N)}_{DW}=E_{DW}(N, g^2) \pm \left(\frac{2}{g^2}\right)^N \frac{ H_0(N, g) \left[\frac{\partial E}{\partial \nu}\right]_{N}}{N!}\,\xi
+ O(\xi^2)
\label{dw-splitting}
\end{eqnarray}

\subsubsection{Resurgent expansion for DW vs. Instantons} 
\label{Sec:res-DW}
We can expand the left hand side of the 
(\ref{dw2})
 up to $k$-th order, and at the same time, the right hand side, up to $(k-1)$-th order in $\delta \nu$. This suffices to systematically extract the trans-series up to  $k$-th order in the instanton expansion. Let us do this exercise for the ground state energy $(N=0)$: Let $\nu = 0+ \delta \nu$, and
 define 
\begin{align}
\sigma_{\pm}= \ln\left(\frac{e^{\pm i\pi}\, 2}{g^2}\right) = \sigma \pm i \pi,  \qquad \sigma = \ln\left(\frac{ 2}{g^2}\right),  
\label{sig-log}
\end{align}
Then, expanding both sides of the global boundary condition we find
\begin{eqnarray}
\frac{1}{\Gamma(-\delta\nu)}\left(\frac{e^{\pm i\pi}\, 2}{g^2}\right)^{-\delta\nu}
&=&  \left\{ -  \delta \nu  Q_0  +  (\delta\nu)^2 Q_1^{\pm} +   (\delta\nu)^3 Q_2^{\pm} +  
\dots\right\} \nonumber\\
&=& -\xi \, H_0(\nu, g^2)  \cr
&=& -\xi \, \left[  H_0 + H_0^{'} (\delta\nu) + \frac{1}{2} H_0^{''} (\delta\nu)^2 +  \ldots \right]
\label{dw3b}
\end{eqnarray}
where $H_0=H_0(0, g^2)$, $H_0^\prime=[\frac{\partial H_0(\nu, g^2)}{\partial \nu}]_{\nu=0}$, etc, and $Q_n(\sigma)$ is an $n$-th order polynomial, encoding the quasi-zero mode integrations in the instanton picture as described below,  the first few of which are given by
\begin{align} 
Q_0 \equiv Q_0(\sigma_{\pm}) &= 1,  \cr 
Q_1^\pm \equiv Q_1(\sigma_{\pm}) &=  \gamma+    \sigma_{\pm}   
\cr
Q_2^\pm \equiv  Q_2(\sigma_{\pm}) &=  - \frac{1}{2} (\gamma+    \sigma_{\pm})^2  + \frac{\pi^2}{12}  
\cr
Q_3^\pm \equiv  Q_3(\sigma_{\pm}) &=   \frac{1}{6} (\gamma+    \sigma_{\pm})^3  - \frac{\pi^2}{12} ( \gamma+    \sigma_{\pm}) - \psi^{(2)}(1) 
\label{poly}
\end{align}
The subscript $n$ counts,  in the instanton picture, the number of quasi-zero modes associated with $(n+1)$-instanton events. 
Solving for $\delta\nu$ iteratively in the instanton fugacity $\xi$, we write  $\delta\nu = \sum a_n \xi^{n}$. Then, it is easy to show that 
 \begin{align} 
\delta\nu &= \xi H_0  \cr
& +  \xi^2  [ H_0 H_0^{'} + H_0^2 Q_1 (\sigma_{\pm}) ]  \cr
&+  \xi^3 \left[ H_0 (H_0^{'})^2 +  \frac{1}{2} H_0^2 H_0^{''} 
  +3  H_0^2H_0^{'}    Q_1 (\sigma_{\pm})    - 3 H_0^3 Q_2  (\sigma_{\pm})   + \frac{\pi^2}{3} H_0^3   \right]   \cr
&+ \ldots
   \label{trans-1-trt} 
\end{align}

{\it Remarks and connection to instanton picture:}

\begin{figure}[htb]
\includegraphics[angle=0, width=0.90\textwidth]{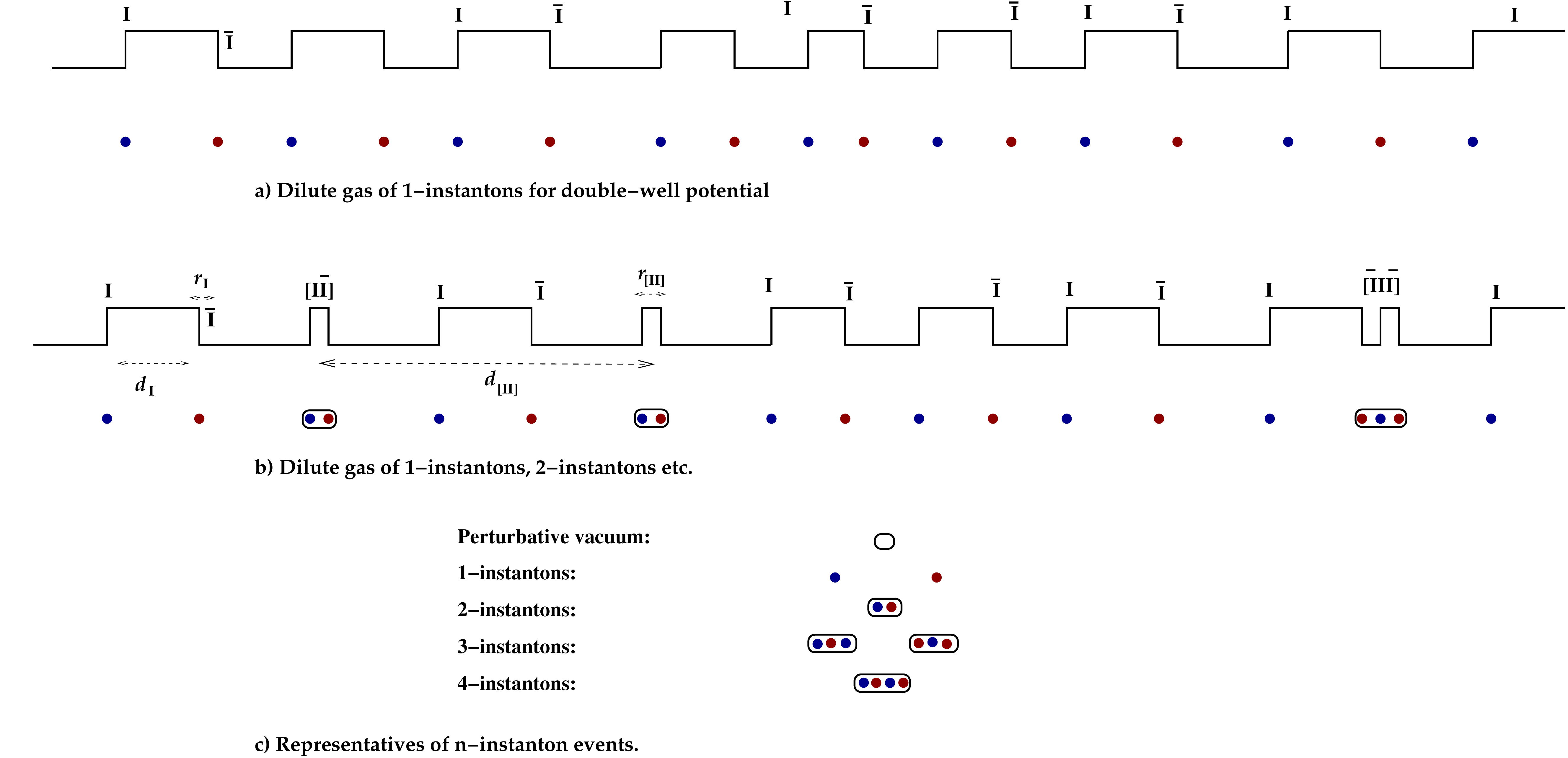}
\caption{  \textsf{ Same as Fig.1, for the double-well potential.}}
\label{moleculesDW-plots}
\end{figure}

We can interpret various terms in the transseries  expansion due to topological  defects with action $n S_I$. 
The   origin of terms proportional to  $\xi$, $\xi^2$, $\xi^3$  are, respectively, 1-,and 2-, and 3-defects, see Fig.\ref{moleculesDW-plots}.
Although there is no strict topological charge, one can still assign a topology to instanton and anti-instanton events by their asymptotics. 
Doing so will help us to disentangle the contributions to physical observable and clarify the cancellations taking place in the (truncated) resurgence triangle.  For ${\cal I}$ and   $\bar {\cal I}$, we assign ``topological charges",  $\pm 1$ as follows:
\begin{eqnarray}
{\cal I}:  \qquad  Q_T = y(\infty) - y(-\infty) = 0 -(-1) = +1  \cr 
\bar {\cal I}:  \qquad Q_T =  y(\infty) - y(-\infty)  = -1 -(0) = -1 
\end{eqnarray} 
Consequently, the topological excitations in the double-well problem and their topological charges are given by 
\begin{eqnarray}
1-{\rm defects}: &&{\cal I} \sim  (\ldots) \xi, \qquad Q_T = +1, \cr\cr  
 &&   \bar {\cal I} \sim  (\ldots) \xi,  \qquad Q_T = -1,    \cr \cr
 2-{\rm defects}: &&  [{\cal I }  \bar {\cal I}]_{\pm} = 
 [\bar {\cal I } {\cal I}]_{\pm} \sim     (\ldots)   \xi^2 \mp i  (\ldots)   \xi^2  \qquad Q_T = 0 
    \cr \cr
 3-{\rm defects}: 
&&    [{\cal I}\bar{\cal I}{\cal I}]_{\pm}  \sim  (\ldots) \xi^3   \pm i (\ldots)\xi^3,     \qquad  Q_T = +1, \cr\cr
   &&  [ \bar{\cal I}{\cal I}\bar{\cal I}]_{\pm} \sim  
   (\ldots) \xi^3   \pm i (\ldots)\xi^3, \qquad  Q_T = -1,    
   \label{inst-trt} 
 \end{eqnarray} 
Note that there are no  $[{\cal I }  {\cal I}],   [ {\cal I}{\cal I}{\cal I}], \ldots   $ type events, i.e, $Q_T \geq 2$ are not present, despite the fact that 
the action of the $n$-event is just  $nS_I$. This is so because we are dealing with a double-well potential. The situation is different for the periodic SG potential: see Section \ref{sg-qs}.
The resurgence triangle \cite{Dunne:2012ae} provides a simple graphical representation of the trans-series structure.
Each cell is labeled by $(n, m)$. 
The rows are sectors with fixed action $(nS_I)$, $n=0, 1,2 ,\ldots $, and  columns  are  sectors with fixed  topological charge $Q_T =m= +1, 0, -1$  (compare with periodic potential, for which  $|m| \leq n$). 
 The truncated resurgence triangle for  the DW-system is 
\begin{eqnarray} 
\underline{m=+1}& \underline{m=0}& \underline{m=-1} \cr
&f_{(0,0)} &\nonumber\\ 
\cr
  e^{-\frac{1}{6g^2} } f_{(1, 1)}  \hskip -20pt
&& \hskip -20pt  \quad
e^{-\frac{1}{6g^2} } f_{(1, -1)} \nonumber\\ \cr
\quad \quad & 
e^{-\frac{2}{6g^2}} f_{(2, 0)}  &
\quad \quad
\nonumber\\   \cr
\quad \quad
e^{-\frac{3}{6 g^2}  } f_{(3,1)}
 \hskip -20pt
&& 
e^{-\frac{3}{6 g^2 }} f_{(3,-1)}
\quad \quad
 \nonumber\\ \cr
\quad \quad
 \quad \quad 
&   e^{-\frac{4}{6 g^2 }} f_{(4,0)}
  & \quad \quad 
\quad \quad
\nonumber\\ \cr
 \vdots \qquad &\vdots & \qquad  \vdots  
\label{triangle-trt}
\end{eqnarray}
Various comments are in order in connection with the transseries (\ref{trans-1-trt}), instanton and multi-instanton amplitudes (\ref{inst-trt}), and the truncated resurgence triangle (\ref{triangle-trt}).
\begin{itemize}
\item{
In the  instanton picture, the interpretation of  $Q_n(\sigma_{\pm})$ is the following. For an $(n+1)$-instanton event,  there are $(n+1)$ low lying modes. In the non-interacting instanton gas picture, these are $(n+1)$-position moduli of these defects.  Including interaction, 
$n$ of these become quasi--zero modes that needs to be integrated exactly,  and one is the ``center of action" of the $(n+1)$-defect. In this way, one obtains the amplitude of the correlated $(n+1)$-instanton event, and  $Q_n(\sigma_{\pm})$  or   $Q_n(\sigma)$ is its prefactor. (See next item). We call $Q_n$ the  {\it quasi-zero mode polynomial}, the degree of polynomial  $n$ counts the number of quasi-zero modes that are integrated over.  \footnote{The correspondence with the notation of Zinn-Justin \cite{ZinnJustin:1981dx}  is the following.  $Q_1(\sigma) = P_2(\sigma)$, $2  (Q_1 )^2+ Q_2 = P_3(\sigma)$, etc.   ZJ uses subscript $(n+1)$ for a polynomial of degree $n$, because the polynomial multiplies an $(n+1)$-event amplitude.  Since the degree $n$ of the polynomial is equal to the number of integrated quasi-zero modes, and the number of QZM is one-less than the number of the constituents of a correlated event, we call this polynomial $Q_n(\sigma)$. 
}}
\item{The polynomials are of two types:  those which are two-fold ambiguous  $Q_n^\pm = Q_n(\sigma^{\pm})$ and those which are not 
$Q_n = Q_n(\sigma)$. 
For the $(n+1)$-instanton  configuration with only instantons,  unambiguous polynomials  $Q_n(\sigma)$ arise.  (This does not happen in the double-well system, because an instanton is always followed by an anti-instanton. But it does happen for the periodic potential as we discuss later.)
Whenever there are both correlated instantons and anti-instantons pairs in an $(n+1)$-instanton event,  polynomials with ambiguities arise. Such is always the case in double-well potential. }
\item{These polynomials are  universal.  They will appear in any QM mechanics problems with degenerate minima.\footnote{They also have natural generalization to QFT, which is not explored here.} 
 In any given theory, we have to consider both instantons as well as 
correlated/molecular $(n+1)$-events.  These polynomials are an   integral part of the $(n+1)$-correlated instanton events. }

\item{The ambiguities in $Q_{n} (\sigma_{\pm} )$ cancel the ambiguities associated with the non-Borel summability of the perturbation theory according to the  rules of resurgence  trianlge. For example, the ambiguity in $Q_1^{\pm}$  and $Q_2^{\pm}$ 
cancel the ambiguities  associated with non-Borel summability of the perturbation theory around perturbative vacuum   and  one-instanton sector, respectively.}
\end{itemize}

{\bf The truncated resurgence  triangle and (graded) partition functions:}
The structure associated with the truncated  resurgence triangle can also be seen by studying  partition functions graded by parity symmetry.  Parity in our DW potential is defined as  
\begin{align}
P: y \rightarrow -1-y  \qquad ({\rm reflection \; w.r.t.  })  \;  y= - \frac{1}{2}
\end{align} 
and commutes with the Hamiltonian, $[P, H]=0$.

  We can define two types of partitions functions, one regular, and one twisted by the insertion of the parity operator: 
  \begin{align}
  Z(\beta, g^2) &= {\rm tr} \;  e^{-\beta H} \longrightarrow  \int_{y(t+ \beta)=  y(t)  } Dy(t) e^{-S[y]}   \qquad \qquad \cr
   \tilde Z (\beta, g^2) & = {\rm tr} \;  P  e^{-\beta H} \longrightarrow  \int_{y(t+ \beta)= P[y(t)]= -y(t)-1  } Dy(t) e^{-S[y]} 
  \end{align}
  The boundary conditions associated with    $Z(\beta, g^2)$ forbids the contribution of a single instanton effect as well as  
  any topological configuration which has the same asymptotic behavior as the single instanton. 
On the flip side, the boundary conditions associated with  
  $\tilde Z (\beta, g^2)$   forbids the contribution of the perturbative vacuum  saddle as well as any topological configuration which has the same asymptotic behavior as the perturbative vacuum.    i.e. $Z(\beta, g^2)$ receives contribution from the   $m= 0$ column, 
  while  $\tilde Z (\beta, g^2)$   receives contribution from the 
  $ \pm 1$ columns in the  resurgence triangle. In the periodic SG potential,  the columns are characterized by a winding number associated with their $\theta$-angle dependence: see Section \ref{sg-qs}.

\subsection{Global Boundary Condition for the Sine-Gordon System}

A similar analysis applies to the SG system. We first take the appropriate linear combinations of the two linearly independent parabolic cylinder functions to match the normalization conditions for the functions $w_I(y)$ and $w_{II}(y)$ in (\ref{basis}). Define even and odd functions on the interval $y\in \left[-\frac{\pi}{2}, +\frac{\pi}{2}\right]$:
\begin{eqnarray}
f_1(y)&=&\frac{1}{\sqrt{u^\prime(y)}}\left(D_\nu\left(\frac{u(y)}{g}\right)+D_\nu\left(-\frac{u(y)}{g}\right)\right) \qquad ({\rm even})
\label{evenf}\\
f_2(y)&=&\frac{1}{\sqrt{u^\prime(y)}}\left(D_\nu\left(\frac{u(y)}{g}\right)-D_\nu\left(-\frac{u(y)}{g}\right)\right) \qquad ({\rm odd})
\label{oddf}
\end{eqnarray}
where we note that $u(y)$ is odd, and $u^\prime(y)$ is even and positive on this interval. The Wronskian is:
\begin{eqnarray}
{\mathcal W}\equiv f_1(y)\, f_2^\prime(y)-f_1^\prime(y)\, f_2(y)=-\frac{4}{g}\sqrt{\frac{\pi}{2}}\frac{1}{\Gamma(-\nu)}
\label{wronskian}
\end{eqnarray}
which is independent of $y$, and is non-zero except when $\nu$ is a non-negative integer. Then, the appropriately normalized basis solutions (\ref{basis}) can be written as:
\begin{eqnarray}
w_{I}(y)&=& \frac{1}{{\mathcal W}}\left(f_2^\prime\left(-\frac{\pi}{2}\right)\, f_1\left(y\right) - f_1^\prime\left(-\frac{\pi}{2}\right)\, f_2(y)\right) 
\label{w1}\\
w_{II}(y)&=& \frac{1}{{\mathcal W}}\left(-f_2\left(-\frac{\pi}{2}\right)\, f_1(y) + f_1\left(-\frac{\pi}{2}\right)\, f_2(y)\right) 
\label{w2}
\end{eqnarray}
Using the parity properties of $f_1$ and $f_2$ we can therefore write the Bloch condition (\ref{bloch}) in various equivalent ways:
\begin{eqnarray}
\cos\theta&=&\frac{1}{{\mathcal W}}\left(f_2^\prime\left(\frac{\pi}{2}\right)\, f_1\left(\frac{\pi}{2}\right) + f_1^\prime\left(\frac{\pi}{2}\right)\, f_2\left(\frac{\pi}{2}\right)\right)
\label{b1}\\
&=&1+\frac{2}{{\mathcal W}}\, f_1^\prime\left(\frac{\pi}{2}\right)\, f_2\left(\frac{\pi}{2}\right)
\label{b2}\\
&=&-1+\frac{2}{{\mathcal W}}\, f_2^\prime\left(\frac{\pi}{2}\right)\, f_1\left(\frac{\pi}{2}\right) 
\label{b3}
\end{eqnarray}
Thus, as in the double-well case, the global boundary condition is imposed at the midpoint between two neighboring perturbative vacua: $y_{\rm midpoint}=\frac{\pi}{2}$. Moreover, the global condition is expressed in terms of  parabolic cylinder functions evaluated at $y_{\rm midpoint}$. This Bloch condition results in the perturbative energy level splitting into a continuous band, with states within the band labeled by the angular parameter $\theta$. The bottom of the lowest band has $\theta=0$ and its wave function is an even function, while the top of the lowest band $\theta=\pi$ and its wave function is an odd function. For these lowest band-edge states the Bloch condition takes a simpler form reminiscent of the DW case (\ref{gdw1}, \ref{gdw2}):
\begin{eqnarray}
({\rm lower, \, even \, state}; \theta=0) &:& f_1^\prime\left(y_{\rm midpoint}\right)=0
\label{sg-lower}\\
({\rm upper, \,odd \, state}; \theta=\pi ) &:& f_1\left(y_{\rm midpoint}\right)=0
\label{sg-upper}
\end{eqnarray}
The Bloch condition determines the ansatz parameter $\nu$ as a function of the coupling $g^2$, for each value of the Bloch angle $\theta$. As before, $u\left(\frac{\pi}{2}\right)$ is a non-Borel-summable divergent series in $g^2$, so we need to analytically continue in $g^2$ off the Stokes line ($g^2>0$) in order to properly define the Bloch condition. This requires again the full analytic continuation behavior (\ref{full}) of the parabolic cylinder functions that enter into the definition of the functions $f_1$ and $f_2$. Proceeding in a manner similar to DW potential, we can write the boundary conditions 
(\ref{b1}, \ref{b2}, \ref{b3}) as an equation determining $\nu$ in terms of $g^2$: 
\begin{eqnarray}
\frac{1}{\Gamma(- \nu)}\left(\frac{ 2}{g^2}\right)^{-\nu}   \pm\, i \frac{\pi}{2} \left(\frac{e^{\pm i\pi}\, 2}{g^2}\right)^{+\nu}  \frac{\xi^2 [H_0 (\nu, g^2)]^2 }{\Gamma(1+ \nu)} = -\xi \, H_0(\nu, g^2)  \cos \theta 
\label{pp3}
\end{eqnarray}
This is  the analog  of the (\ref{dw2}) for DW, now applied to  SG potential.

At leading non-perturbative order, we find that the parameter $\nu$ is exponentially close to an integer. For example, for the lowest band we write $\nu=0+\delta\nu+\dots$, and the Bloch condition  (\ref{b1})  becomes in the small $g^2$ limit:
\begin{eqnarray}
\cos\theta &=& -\frac{g}{4}\sqrt{\frac{2}{\pi}}\,\Gamma(-\nu)
\left(f_2^\prime\left(\frac{\pi}{2}\right)\, f_1\left(\frac{\pi}{2}\right) + f_1^\prime\left(\frac{\pi}{2}\right)\, f_2\left(\frac{\pi}{2}\right)\right)
\label{b4}\\
&\sim & \frac{g}{4}\sqrt{\frac{2}{\pi}}\, (\delta\nu)\, \frac{\pi}{2u(\pi/2)}\exp\left[\frac{(u(\pi/2))^2}{2g^2}\right]
\end{eqnarray}
Using from (\ref{upsg}) the fact that
\begin{eqnarray}
u\left(\frac{\pi}{2}\right)\sim 2-\frac{g^2}{4}(2\nu+1)\ln 2+\dots 
\label{umid}
\end{eqnarray}
we find that 
\begin{eqnarray}
\delta \nu\sim -\frac{4}{\sqrt{\pi}\, g}\, \cos\theta \, e^{-\frac{2}{g^2}}
\label{dnu}
\end{eqnarray}
which gives the familiar instanton result for the splitting of the lowest band. Incorporating fluctuation terms we find:
\begin{eqnarray}
E_\theta^{({\rm lowest \, band})}\sim \left[1-\frac{g^2}{4}-\frac{g^4}{16}-\dots\right]- \cos\theta \,\frac{8}{\sqrt{\pi}\, g}\, e^{-\frac{2}{g^2}}\left[1-\frac{7g^2 }{16}-\frac{59 g^4}{512} -\dots\right]+O\left(e^{-\frac{4}{g^2}}\right) \qquad
\label{sg-band}
\end{eqnarray}
This is in agreement with the Mathieu equation results \cite{nist,goldstein,Stone:1977au}.
\begin{figure}[htb]
\includegraphics[scale=0.6]{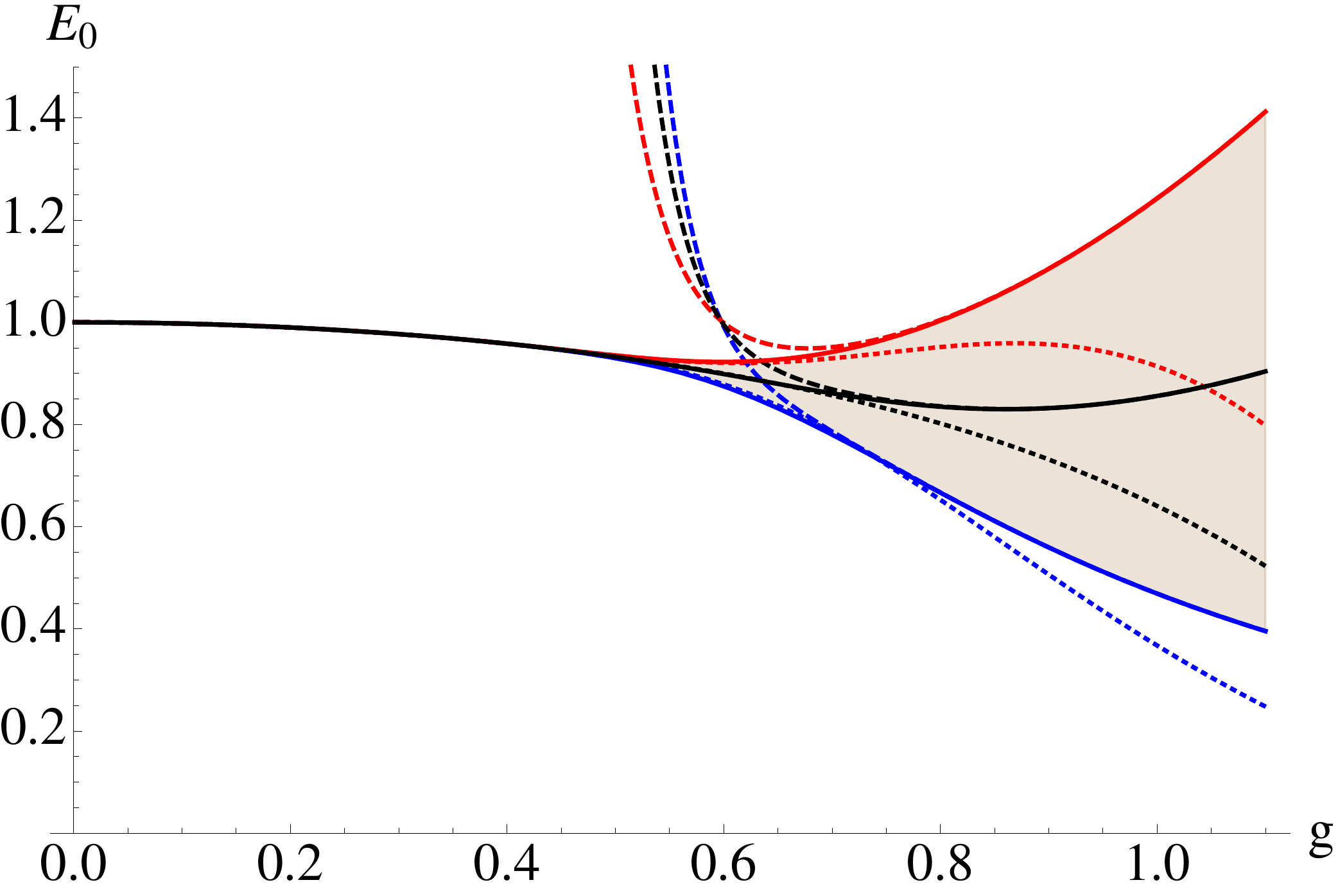}
\caption{\textsf{  A comparison of the exact band edges, and the center of the band, for the lowest band of the Sine-Gordon potential [solid lines], with the weak-coupling trans-series expansion [dotted lines], and the strong-coupling results [dashed lines]. The exact results are generated using the Mathematica functions MathieuCharacteristicA and MathieuCharacteristicB, which compute Mathieu band edges numerically. The weak-coupling expansions have been plotted here using the expression  in (\ref{weak}), and the strong-coupling expansions have been plotted using (\ref{strong}).  Note the excellent numerical agreement.}}
\label{sg-plot}
\end{figure}
In Figure \ref{sg-plot} we show that this expansion gives an excellent approximation to the lowest band for the Sine-Gordon potential. The edges of the lowest band are given by $\theta=0, \pi$ in the small coupling limit, $g^2\to 0$. We can also compute the strong coupling limit, $g^2\to\infty$, by treating the potential as a perturbation of the free periodic or anti-periodic solution on the single-period interval. We obtain the following weak and strong coupling expressions:
\begin{eqnarray}
\left[E_0^\pm\right]_{({\rm weak-coupling})}&=&\left[1-\frac{g^2}{4}-\frac{g^4}{16}-\frac{3g^6}{64}-\dots\right] \pm \frac{8}{\sqrt{\pi} g}e^{-2/g^2}\left[1-\frac{7g^2 }{16}-\frac{59 g^4}{512} -\dots\right] +\dots 
\label{weak}\\
\left[E_0^\pm\right]_{({\rm strong-coupling})}&=&
\begin{cases}
g^2+\frac{1}{4g^2}-\frac{1}{128 g^6}+\dots \cr
\frac{1}{2g^2}-\frac{1}{32g^6}+\frac{7}{2^{15} g^{14}}-\dots 
\end{cases}
\label{strong}
\end{eqnarray}
Note that the strong-coupling expansion is convergent (it is not unusual for functions to have convergent expansions for large/small argument, but asymptotic expansions for small/large argument).
We can also plot the {\it exact} expressions for the band edges, by writing the global boundary conditions (\ref{sg-lower}, \ref{sg-upper}) directly in terms of Mathieu functions, which can be plotted. Figure \ref{sg-plot} shows excellent agreement of the exact result with the asymptotic limits. 

If we include more exponentially small   terms  (along with the perturbative fluctuations around them) in the weak  coupling transseries expansion, the trans-series will  approach to the exact result even for larger values of the coupling.  Analogously, 
if we include more terms in the strong coupling expansion, the series will actually approach to the exact result even for smaller values of the coupling.

\subsubsection{Resurgent expansion for SG vs. Instantons} 
\label{sg-qs}
Similar to Section \ref{Sec:res-DW}, 
we can expand the first(second) term in  left hand side of the (\ref{pp3})
 up to $k$-th  $(k-2)$-th    order,  and at the same time, the right hand side, up to $(k-1)$-th order in $\delta \nu$. 
 This suffices to systematically extract the trans-series up to  $k$-th order in the instanton expansion. 
 Doing so also  helps us to  visualize the differences and similarities  between the resurgent expansions in periodic and double-well potentials.

 For the ground state, we let $\nu= N + \delta \nu = 0 + \delta \nu$. Then, 
\begin{eqnarray}
 \left\{ - \delta \nu Q_0 +  (\delta\nu)^2 Q_1 +   (\delta\nu)^3 Q_2 \right\}     \pm   \frac{ i \pi \xi^2 }{2} \left[  H_0^2 +   (\delta\nu)  ( 2 H_0 H_0' +  Q_1^{\pm} )  \right]   \cr
= \xi \, \left[  H_0 + (\delta\nu)  H_0^{'} + \frac{1}{2} (\delta\nu)^2 H_0^{''}   \right] \cos \theta \qquad 
\label{sg3}
\end{eqnarray}
where $Q_n$  and $Q_n^{\pm}$ are the  quasi-zero mode polynomials given in (\ref{poly}). 
Solving for $\delta\nu$ iteratively in instanton fugacity $\xi$,  we find 
\begin{align} 
\delta\nu =&  - \xi H_0  \cos \theta  \cr
& +  \xi^2 \Big(  \Big[ H_0 H_0^{'} + H_0^2 Q_1 \Big] \cos^2 \theta  \mp \frac{i \pi}{2} H_0^2 \Big)   \cr
&+  \xi^3  \left(  \left[ - H_0 (H_0^{'})^2  -3 H_0^{'} H_0^2 Q_1   -  H_0^3 (2  (Q_1 )^2+ Q_2)     -    \frac{1}{2}H_0^2 H_0^{''}   \right]  \cos^3 \theta \right.
\cr
&\left.  +   \left[  \pm i \pi Q_1 H_0^3 \pm   \frac{3}{2} i \pi H_0^2 H_0'  \pm \frac{1}{2} i \pi H_0^3 Q_1^{\pm}     \right]   \cos \theta \right)  
 + \ldots \cr
=&  - \frac{1}{2} \xi H_0  e^{i \theta}    - \frac{1}{2} \xi H_0  e^{-i \theta}    \cr&+   \frac{1}{4} \xi^2  \Big[ H_0 H_0^{'} + H_0^2 Q_1 \Big]   e^{2 i \theta} +   \frac{1}{2} \xi^2  \Big[ H_0 H_0^{'} + H_0^2 Q_1^{\pm} \Big] +    \frac{1}{4} \xi^2  \Big[ H_0 H_0^{'} + H_0^2 Q_1 \Big]   e^{-2 i \theta} 
+ \ldots \qquad
\label{trans-1}
\end{align}
{\it Remarks and connection to instanton picture:}
We can interpret the various terms in the transseries  expansion due to topological  defects with action $n S_I$ and winding number $m\leq n$. 
For example, the origin of terms proportional to  $\xi$, $\xi^2$, $\xi^3$  are, respectively, 1-,and 2-, and 3-defects. 
Let us classify the $n$-defects contributing to the transseries expansion at order $n$: 
\begin{eqnarray}
1-{\rm defects}: &&{\cal I} \sim  (\ldots) \xi e^{i \theta} ,  \qquad  \bar {\cal I} \sim  (\ldots) \xi e^{-i \theta},  \cr \cr
 2-{\rm defects}: &&
 [ {\cal I}{\cal I}] \sim  (\ldots) \xi^2 e^{2i \theta},    \qquad  [ \bar {\cal I}   \bar  {\cal I}] \sim  (\ldots) \xi^2 e^{-2i \theta}  \qquad  [{\cal I }  \bar {\cal I}]_{\pm} = 
 [\bar {\cal I } {\cal I}]_{\pm} \sim     (\ldots)   \xi^2 \mp i  (\ldots)   \xi^2    \cr \cr
 3-{\rm defects}: && 
 [ {\cal I}{\cal I}{\cal I}] \sim  (\ldots) \xi^3 e^{3i \theta},    \qquad  [ \bar {\cal I}   \bar {\cal I}  \bar  {\cal I}] \sim  (\ldots) \xi^3 e^{-3i \theta},   \qquad  \cr \cr
&&  [ \bar{\cal I}{\cal I}{\cal I}]_{\pm} =  [{\cal I}\bar{\cal I}{\cal I}]_{\pm} = [ {\cal I}{\cal I}\bar{\cal I}]_{\pm} \sim  (\ldots) \xi^3 e^{i \theta}  \pm i (\ldots)\xi^3 e^{i \theta},    \qquad  \cr\cr
   &&  [ \bar{\cal I}\bar{\cal I}{\cal I}]_{\pm} =  [{\cal I}\bar{\cal I}\bar{\cal I}]_{\pm} = [ \bar{\cal I}{\cal I}\bar{\cal I}]_{\pm} \sim  
   (\ldots) \xi^3 e^{-i \theta}  \pm i (\ldots)\xi^3 e^{-i \theta}     
   \label{inst-1} 
 \end{eqnarray} 
Note the multiplicity of the $n$-defects. At action level $n$, the events with topological charge $m= n-2k,  \; k =0,1, \ldots, n$ has multiplicity  
${n\choose k}$.    For example, 2- and 3-defects  have multiplicities 1,2,1 and 
$1,3,3,1$, respectively, and combine to give  $\cos^2 \theta, \cos^3 \theta$ terms in the transseries  (\ref{trans-1}).
 In general,  we have instanton events of the form $[{\cal I}^{n-k} \bar{\cal I}^k]$ with $n$ units of action and $\theta$ dependence  $e^{i (n-2k) \theta}$.  The multiplicities of this correlated events are  ${n\choose k}$. Hence,  
 \begin{align} \sum_{k=0}^{n} {n\choose k} e^{i (n-2k) \theta}   = \left(  e^{i \theta} + e^{-i \theta}  \right)^n = 2^n   \cos^n \theta\end{align} giving the result obtained above. e.g., in  (\ref{trans-1}). 

A way to organize trans-series is through the structure of the resurgence triangle,   where each cell is labeled by $(n, m), |m| \leq n$. 
The rows are sectors with fixed action $(nS_I)$, $n=0, 1,2 ,\ldots $, and  columns  are  sectors with fixed  topological charge $|m|  \leq n$.    The resurgence triangle for  the periodic potential is: 
\begin{eqnarray}
&f_{(0,0)} &\nonumber\\ 
\cr
  e^{-\frac{2}{g^2}+i  \theta } f_{(1, 1)}  \hskip -20pt
&& \hskip -20pt  \quad
e^{-\frac{2}{g^2} -i \theta } f_{(1, -1)} \nonumber\\ \cr
e^{-\frac{4}{g^2}+2i \theta  } f_{(2, 2)}
\quad \quad & 
e^{-\frac{4}{g^2}} f_{(2, 0)}  &
\quad \quad
e^{-\frac{4}{g^2}-2i  \theta} f_{(2, -2)}
\nonumber\\   \cr
e^{-\frac{6}{g^2}+3i   \theta} f_{(3,3)}
\quad \quad
e^{-\frac{6}{g^2}+i   \theta  } f_{(3,1)}
 \hskip -20pt
&& 
e^{-\frac{6}{g^2}-i  \theta} f_{(3,-1)}
\quad \quad
e^{-\frac{6}{g^2}-3i   \theta} f_{(3, -3)} \nonumber\\ \cr
e^{-\frac{8}{g^2}+4i  \theta } f_{(4,4)}
\quad \quad
e^{-\frac{8}{g^2}+2i  \theta  } f_{(4,2)} \quad \quad 
&   e^{-\frac{8}{g^2} } f_{(4,0)}
  & \quad \quad 
e^{-\frac{8}{g^2}-2i  \theta} f_{(4,-2)}
\quad \quad
e^{-\frac{8}{g^2}-4i   \theta } f_{(4, -4)} \nonumber\\ \cr
\iddots \qquad \qquad \qquad \qquad\qquad &\vdots & \qquad \qquad\qquad \qquad\qquad  \ddots 
\label{triangle}
\end{eqnarray}
Various comments are in order in connection with the transseries (\ref{trans-1}), instanton and multi-instanton amplitudes (\ref{inst-1}), and the resurgence triangle (\ref{triangle}).

 \begin{itemize} 
\item The $n$-instanton, e.g.,  $  {\cal I}, \; [ {\cal I}{\cal I}], \;   [ {\cal I} {\cal I}\ldots {\cal I}] $  (similarly for $n$-anti-instanton) amplitudes, associated with the edges of triangle  $m=\pm n$  
are the leading semi-classical configuration in the corresponding homotopy class, and  
unambiguous.  This is because instanton-instanton interactions are repulsive and defining the $n$-instanton amplitude does not require the  BZJ-prescription.   The quasi-zero mode integrations produce ambiguity-free $Q_{n-1} (\sigma)$ polynomials.  
 However, the perturbative expansion around the  $n$-instanton  is  non-Borel summable and still ambiguous.  

\item Since $n$-instanton amplitude has $\theta$-dependence $e^{i n \theta}$, it cannot mix with perturbation theory around the perturbative vacuum, which is clearly insensitive to $\theta$.   
 Since  the basis \{$e^{i n \theta}, \; n\in \mathbb Z$\} forms an orthogonal complete set for periodic functions with period $2\pi$,  we can define superselection sectors in the resurgence triangle, i.e.,  columns  with different $\theta$ angle dependence  are associated with different homotopy classes and  do not mix with each other in the cancellation of their ambiguities.    

\item
The ambiguous  part in the $O(\xi^2)$ term does not depend on the Bloch angle $\theta$. 
 The contribution of  $[\bar {\cal I } {\cal I}]_{\pm}$  produce ambiguous 
quasi-zero mode polynomials  $Q_{1}^{\pm} = Q_{1} (\sigma_{\pm})$.  
 It must be this way if this imaginary  term is to cancel against an imaginary term arising from the non-Borel-summable perturbative series, because the perturbative series is  independent of $\theta$. This pattern continues throughout the entire trans-series, and the  resurgence triangle \cite{Dunne:2012ae}, in which the resurgent cancellations are characterized by their $\theta$ dependence.
  
\end{itemize}

{\bf The resurgence  triangle and graded partition functions:}
The structure associated with the resurgence triangle can also be seen by studying  graded partition functions.  
Consider the Fourier 
expansion of the partition function in the orthonormal basis  \{$e^{i n \theta}, \; n\in \mathbb Z$\}: 
\begin{align}
Z(\beta, g, \theta)= \sum_{m=-\infty}^{+ \infty} e^{i m \theta} Z_{m} (\beta, g) 
\end{align}
Then, it is not hard to realize that $Z_{m} (\beta, g)$ has a resurgent expansion associated with with $m$-th column in  (\ref{triangle}).  In the operator formalism, this data can be extracted by studying the  twisted partition function $Z_{m} ={\rm tr} \; T^{ m} e^{-\beta H} $ with the insertion of the translation operator $T$. In path integrals, this corresponds to restricting the boundaries of the path integration, namely:
\begin{align}
Z_{m} ={\rm tr} \; T^{ m} e^{-\beta H} \longrightarrow  \int_{x(t+ \beta)=  x(t) +  m \frac{\pi}{g} } Dx(t) e^{-S[x]} 
\end{align}
For example, the boundary conditions associated with  the $Z_{\pm 1}$  {\bf  forbids} the contribution of the perturbative vacuum sector, as well as any topological configuration which has the same asymptotic behavior as the perturbative vacuum. The leading saddle contributing to $Z_{\pm 1}$   is 
a one-instanton event, from which one can extract the band-width at leading order.

\section{Explicit Resurgence Relations}
\label{sec:resurgence}

\subsection{Comparison with Zinn-Justin and Jentschura}

In order to discuss the explicit resurgent relations encoded in the trans-series expressions for the energy eigenvalues, it is convenient at this stage to comment on the similarities and differences between the uniform WKB approach, discussed in this paper, and the approach of Zinn-Justin and Jentschura (ZJJ), who have presented extensive results for the resurgent relations \cite{zjj}. 

\subsubsection{Double-well Potential}

For the DW problem,  ZJJ express their exact quantization condition as \cite{zjj}
\begin{eqnarray}
\frac{1}{\sqrt{2\pi}}\Gamma\left(\frac{1}{2}-B(E, g^2)\right)\left(\frac{2}{g^2}\right)^{B(E, g^2)} e^{-A(E,  g^2)/2}=\pm i
\label{zjj-dw}
\end{eqnarray}
where the $\pm$ sign in (\ref{zjj-dw}) refers to the splitting of a given perturbative level into two separate levels, and the perturbative function $B(E, g^2)$ and the non-perturbative function $A(E, g^2)$ were computed to be (converting the results of \cite{zjj} to our notation: $E_{ZJ}\to \frac{E}{2}$, and $g_{ZJ}\to g^2$): 
\begin{eqnarray}
B_{\rm DW}(E, g^2)&=&\frac{E}{2}+g^2\left(\frac{3}{4}E^2+\frac{1}{4}\right)
+g^4\left(\frac{35}{8} E^3+\frac{25}{8}E\right)
+g^6\left(\frac{1155}{32}E^4+\frac{735}{16}E^2+\frac{175}{32}\right)\nonumber\\
&&
+g^8\left(\frac{45045}{128}E^5+\frac{45045}{64}E^3+\frac{31185}{128}E\right)+\dots
\label{bdw}
\end{eqnarray}
\begin{eqnarray}
A_{\rm DW}(E, g^2)&=&\frac{1}{3g^2}+g^2\left(\frac{17}{4}E^2+\frac{19}{12}\right)
+g^4\left(\frac{227}{8} E^3+\frac{187}{8}E\right)
+g^6\left(\frac{47431}{192}E^4+\frac{34121}{96}E^2+\frac{28829}{576}\right)\nonumber\\
&&
+g^8\left(\frac{317629}{128}E^5+\frac{264725}{48}E^3+\frac{842909}{384}E\right)+\dots
\label{adw}
\end{eqnarray}
Notice that our global boundary condition (\ref{dw2}) has the form
\begin{eqnarray}
\frac{1}{\sqrt{2\pi}}\Gamma\left(\frac{1}{2}-B\right)\left(\frac{2}{g^2}\right)^{B} e^{-A(B,  g^2)/2}=\pm i
\label{gdw}
\end{eqnarray}
where $B\equiv \nu+\frac{1}{2}$, and $A=A(B, g^2)$ is a known function of $B$ and $g$, given in (\ref{dw2}, \ref{xidw}, \ref{xih}). 

To understand the precise relation between ZJJ's result (\ref{zjj-dw}-\ref{adw}) and our expression (\ref{gdw}), observe that if we invert the expression (\ref{bdw}) for $B=B(E, g^2)$ to write it as $E=E(B, g^2)$ we obtain 
\begin{eqnarray}
E_{\rm DW}(B, g^2)&=&2B-2g^2\left(3B^2+\frac{1}{4}\right)-2g^4\left(17 B^3+\frac{19}{4}B\right)-2g^6\left(\frac{375}{2}B^4+\frac{459}{4}B^2+\frac{131}{32}\right)\nonumber\\
&&
-2g^8\left(\frac{10689}{4}B^5+\frac{23405}{8}B^3+\frac{22709}{64}B\right)-\dots
\label{ebdw1}
\end{eqnarray}
which agrees precisely the perturbative expansion (\ref{ebdw0}) for $E(B, g^2)$ that was found in the perturbative expansion of the uniform WKB approach. Recall that this is exactly the usual perturbative expansion for the energy of the $N^{\rm th}$ level, when we identify $B=N+\frac{1}{2}$.
Moreover, if we now insert this expression for $E=E_{\rm DW}(B, g^2)$ as a function of $B$ into ZJJ's expression (\ref{adw}) for $A=A_{\rm DW}(E, g^2)$, we obtain 
the expansion of $A_{\rm DW}(B, g^2)$ in powers of the coupling:
\begin{eqnarray}
A_{\rm DW}(B, g^2)&=&\frac{1}{3g^2}+g^2\left(17B^2+\frac{19}{12}\right)+g^4\left(125 B^3+\frac{153}{4}B\right)+g^6\left(\frac{17815}{12}B^4+\frac{23405}{24}B^2+\frac{22709}{576}\right)\nonumber\\
&&
+g^8\left(\frac{87549}{4}B^5+\frac{50715}{2}B^3+\frac{217663}{64}B\right)+ \dots
\label{adw2}
\end{eqnarray}
This matches precisely the function $A_{\rm DW}(B, g^2)$ obtained from our global condition (\ref{dw2}, \ref{xidw}, \ref{xih}). 

Thus, the conditions (\ref{zjj-dw}) and (\ref{gdw}) are equivalent. However, the philosophy is subtly different. In \cite{zjj}, the expression (\ref{zjj-dw}) is regarded as an equation for the energy $E$ as a function of $g^2$, provided both functions $B(E, g^2)$ and  $A(E, g^2)$ are known. On the other hand, we regard  (\ref{gdw}) as an equation for $B$ (equivalently for $\nu\equiv B-1/2$) as a function of $g^2$, provided the  function $A(B, g^2)$ is known, and we then insert the resulting $B(g^2)$ into the perturbative expansion (\ref{pte}) in order to obtain the resurgent trans-series expression for the energy eigenvalue. We will see in Section \ref{sec:explicit} that there is a surprising advantage to the latter, uniform WKB, perspective.

\subsubsection{Sine-Gordon Potential}

A similar correspondence applies to the SG potential. ZJJ express their exact (Bloch) quantization condition as
\begin{eqnarray}
\left(\frac{2}{g^2}\right)^{-B(E, g^2)}  \frac{e^{A(E, g^2)/2}}{\Gamma\left(\frac{1}{2}-B(E, g^2)\right) }+
\left(-\frac{2}{g^2}\right)^{B(E, g^2)}  \frac{e^{-A(E, g^2)/2}}{\Gamma\left(\frac{1}{2}+B(E, g^2)\right) }=\frac{2\, \cos\theta}{\sqrt{2\pi}}
\label{zjj-sg}
\end{eqnarray}
where  $\theta$ is the Bloch angle, and the perturbative function $B(E, g^2)$ and the non-perturbative function $A(E, g^2)$ were computed to be (converting the results of \cite{zjj} to our notation: $E_{ZJ}\to \frac{E}{2}$, and $g_{ZJ}\to \frac{g^2}{4}$): 
\begin{eqnarray}
B_{\rm SG}(E, g^2)&=&\frac{1}{2}E+\frac{g^2}{16}\left(1+E^2\right) 
+\frac{g^4}{128} \left(5E+3 E^3\right) 
+\frac{g^6}{64}\left(\frac{17}{32}+\frac{35}{16}E^2 +\frac{25}{32}E^4\right) \nonumber\\
&&+\frac{g^8}{256}\left(\frac{721}{128}E+\frac{525}{64}E^3+\frac{245}{128} E^5\right) +\dots
\label{bsg0}
\end{eqnarray}
\begin{eqnarray}
A_{\rm SG}(E, g^2)&=&\frac{4}{g^2}+\frac{3g^2}{16}\left(1+E^2\right)
+\frac{g^4}{16} \left(\frac{23}{4} E+\frac{11}{8} E^3\right) 
+\frac{g^6}{64} \left(\frac{215}{64}+\frac{341}{32}E^2 +\frac{199}{64} E^4\right)\nonumber\\
   &&+\frac{g^8}{256} \left(\frac{4487}{128} E+\frac{326}{8} E^3+\frac{1021}{128} E^5\right)+\dots
\label{asg}
\end{eqnarray}
(Note there is a small typo in (6.32) of \cite{zjj}. The term $-\frac{199}{4}$ should be $+\frac{199}{4}$).

We invert the first expression (\ref{bsg0}) to obtain
\begin{eqnarray}
E_{\rm SG}(B, g^2)&=& 2B-\frac{g^2}{2}\left(B^2+\frac{1}{4}\right)
-\frac{g^4}{8}\left(B^3+\frac{3}{4} B\right) 
-\frac{g^6}{32}\left(\frac{5}{2} B^4+\frac{17}{4} B^2+\frac{9}{32}\right) \nonumber\\
   &&-\frac{g^8}{128}\left(\frac{33}{4} B^5+\frac{205}{8} B^3+\frac{405}{64} B\right) -\dots
\label{bsg2}
\end{eqnarray}
which agrees precisely with the perturbative expression (\ref{ebsg}) found in the uniform WKB approach.
Substituting $E_{\rm SG}(B, g^2)$ for $E$ in order to re-express $A$ as $A=A(B, g^2)$, we find:  
\begin{eqnarray}
A_{\rm SG}(B, g^2)&=&\frac{4}{g^2}+\frac{g^2}{4}\left(3 B^2+\frac{3}{4}\right) 
+\frac{g^4}{16}\left(5 B^3+\frac{17}{4} B\right)
 +\frac{5g^6}{4096} \left(176 B^4+328 B^2+27\right) \nonumber\\
   &&+\frac{9 g^8}{16384} \left(336 B^5+1120 B^3+327 B\right)    +\dots
 \label{asg2}
\end{eqnarray}
In the ZJJ approach \cite{zjj}, the  expression (\ref{zjj-sg}) determines the energy $E$ as a function of $g^2$, provided both functions $B(E, g^2)$ and $A(E, g^2)$ are known. On the other hand, in the uniform WKB approach, this same condition is viewed as determining $B$ as a function of $g^2$, given the function $A(B, g^2)$, and this is then inserted into the perturbative expansion $E(B, g^2)$ to determine the energy.

\subsection{Cancellation of Ambiguities (beyond  Bogomolny--Zinn-Justin)}
\label{sec:cancel}


We first demonstrate the cancellation between the ambiguous imaginary terms arising from the non-Borel-summability of the perturbative series and the ambiguous imaginary terms arising from the analytic continuation in $g^2$ in the 
global boundary condition including  perturbative fluctuations around the non-perturbative 
factors.  This cancellation of ambiguities at two-instanton order  is known as Bogomolny--Zinn-Justin 
mechanism.  Below, we provide evidence that this is also true if one includes perturbative  fluctuations around the non-perturbative saddle $[{\cal I} \bar {\cal I}]$. 
\subsubsection{Double-Well Potential}
The energy transseries for the level  $N$ can be written as  
\begin{eqnarray}E^{(N)}(g^2)=E(B=N+\frac{1}{2}, g^2)+\delta \nu\, \left[\frac{\partial E(B, g^2)}{\partial B}\right]_{B=N+\frac{1}{2}}+\frac{1}{2}(\delta \nu)^2\, \left[\frac{\partial^2 E(B, g^2)}{\partial B^2}\right]_{B=N+\frac{1}{2}}+ \dots
\end{eqnarray}
The first term is the perturbative series, which is non-Borel-summable. The resummation results in the imaginary ambiguous term of order two-instantons, $e^{-2S_I/g^2}$. 
For the reality of the resurgent transseries for real coupling, this must be canceled by an imaginary part coming from the higher non-perturbative (NP) terms in the trans-series.

From (\ref{nu-exp}) we see that the first imaginary term arises in the $O(\xi^2)$ term in $\delta\nu$, which is the two-instanton sector. To this order it has the form
\begin{eqnarray}
{\rm Im}(\delta \nu)&=&\pm \pi \left(\frac{\left(-\frac{2}{g^2}\right)^N}{N!}\exp\left[-\frac{1}{2}\left(A_{\rm DW}\left(N+\frac{1}{2}\right) -\frac{1}{3g^2}\right)\right]\right)^2 \xi^2+\dots \\
&=&
\pm \pi 
\left(\frac{\left(-\frac{2}{g^2}\right)^{N}}{N!}\right)^2 \left[1-g^2\, q_2\left(N+\frac{1}{2}\right)+g^4\left(\frac{1}{2}q_2^2\left(N+\frac{1}{2}\right)-q_3\left(N+\frac{1}{2}\right)\right) +\dots\right]\xi^2+\dots
\label{nua}
\end{eqnarray}
where the polynomials $q_k(B)$ are defined in  (\ref{adw2}):
\begin{eqnarray}
A_{\rm DW}(B, g)-\frac{1}{3g^2}\equiv \sum_{k=1}^\infty g^{2k} q_{k+1}(B)
\end{eqnarray}
Note that the prefactor of $\xi^2$ is a perturbative series  in $g^2$.

The leading  imaginary part  of the energy coming from the two-instanton sector, including the perturbative fluctuations around it, can be found by calculating ${\rm Im}\left( \left[\delta \nu\frac{\partial E}{\partial B}\right]_{B=\frac{1}{2}}\right)$. 
Using (\ref{ebdw1}) we find:
\begin{eqnarray}
\frac{\partial E_{\rm DW}}{\partial B}=2\left(1-6 B g^2-\left(51B^2+\frac{19}{4}\right)g^4+\dots\right)
\end{eqnarray}
For example, for the $N=0$ level we get  (recall, $\xi^2 \sim e^{-2S_I/g^2}$)
\begin{eqnarray}
{\rm Im}\left( \left[\delta \nu\frac{\partial E}{\partial B}\right]_{B=\frac{1}{2}}\right)
&=&\pm 2 \pi \left(1-\frac{35 g^2}{6}-\frac{1277 g^4}{72}\right) \left(1-3g^2-\frac{35 g^4}{2}\right) \xi^2\\
&=&\pm 2\pi 
\left(1-\frac{53}{6} g^2-\frac{1277}{72}g^4-\dots\right)\xi^2
\end{eqnarray}
Compare this with the large-order behavior of perturbation theory quoted  in eqn (8.7) of \cite{zjj} (converted to our notation)
\begin{eqnarray}
E^{(0)}_k &\sim& 
 -\frac{2}{\pi}\, 3^{k+1}k! \left[1-\frac{53}{6}\frac{1}{(3k)}-\frac{1277}{72}\frac{1}{(3k)^2} -\dots \right]
 \label{fluc-dis}
\end{eqnarray}
 Given the subleading corrections to large-order terms (\ref{fluc-dis}), we can obtain 
 the imaginary part by standard dispersion relation arguments \cite{zinnbook}.
Remarkably, not only does the leading term cancel, but also the sub-leading terms are canceled once we include the prefactor. This precise correspondence between the coefficients of the behavior of high orders of perturbation theory about the vacuum and the coefficients of the low orders of fluctuations about the 2-instanton sector is an explicit example of ``resurgence''. 
The behavior near one saddle  (P-saddle) ``resurges'' in the behavior near another saddle  ($[{\cal I} \bar {\cal I}]$, an NP-saddle)
 \cite{Ecalle:1981}.
 

\subsubsection{Sine-Gordon Potential}

For the SG potential, we note the important distinction that the imaginary part in the $O(\xi^2)$ term does not depend on the Bloch angle $\theta$. It must be this way if this term is to cancel against an imaginary term arising from the non-Borel-summable perturbative series, because the perturbative series is clearly independent of $\theta$. 
For example, for the $N=0$ level we get

\begin{eqnarray}
{\rm Im}\left( \left[\delta \nu\frac{\partial E_{\rm SG}}{\partial B}\right]_{B=\frac{1}{2}}\right)
&=&\pm 2 \pi \left(1-\frac{3 g^2}{8}-\frac{13 g^4}{128}\right) \left(1-\frac{g^2}{4}-\frac{3 g^4}{32}\right) \xi^2\\
&=&\pm 2\pi 
\left(1-\frac{5}{2} \frac{g^2}{4}-\frac{13}{8}\left(\frac{g^2}{4}\right)^2-\dots\right)\xi^2
\end{eqnarray}
Compare this with the large-order behavior of perturbation theory quoted  in Appendix A  of 
\cite{Stone:1977au} (converted to our notation) 
\begin{eqnarray}
E^{(0)}_k &\sim& 
 -\frac{2}{\pi}\,k! \left[1-\frac{5}{2k}-\frac{13}{8}\frac{1}{k^2} -\dots \right]
\end{eqnarray}
from which we obtain the imaginary part by standard dispersion relation arguments \cite{zinnbook}.

To recap,  for the DW and SG problems,  the instanton actions are given by
$S_I^{\rm DW}= \frac{1}{6}$ and $ S_I^{\rm SG}= \frac{1}{2}$,
respectively.  The instanton factor is
 $\xi =   \sqrt {\frac{2S_I}{\pi g^2}}  e^{-S_I/g^2}$, while the
imaginary part associated with Borel resummation of vacuum  energy is
 $ \pm 2 \pi  \xi^2 =  \pm  2 \times  \frac{2S_I}{g^2}
e^{-2S_I/g^2}$.  Including fluctuations around the
  $ [{\mathcal I} \bar {\mathcal I}]$  in the transseries, we can write
\begin{align}
 {\rm Im} \left(2 \times  [{\mathcal I} \bar {\mathcal I}]_{\pm}
\sum_{k=0}^{\infty} a_k^{[{\mathcal I} \bar {\mathcal I}]}  g^{2k} \right)
 = \pm    \frac{4S_I}{g^2}   e^{-2S_I/g^2} \left(a_0^{[{\mathcal I} \bar
{\mathcal I}]}  +  a_1^{[{\mathcal I} \bar {\mathcal I}]}  g^{2} +
 a_2^{[{\mathcal I} \bar {\mathcal I}]}  g^{4} + \ldots \right) + O( e^{-4S_I/g^2} )
 \end{align}
This   implies that, using the dispersion relations,
\begin{align}
E_k^{0} = \frac{1}{\pi} \int_0^{\infty} {\rm Im} E_0 (g^2)
\frac{d(g^2)}{(g^2)^{k+1}}
\end{align}
 the large order behavior of the perturbation theory (including the
subleading $1/k$ suppressed  terms) is given by
 \begin{eqnarray}
E^{(0)}_k &\sim&
 -\frac{2}{\pi}\, \frac{k!}{(2S_I)^{k}}  \left[ a_0^{[{\mathcal I} \bar
{\mathcal I}]}  + a_1^{[{\mathcal I} \bar {\mathcal I}]} \left( \frac{2S_I}{k}
\right)+  a_2^{[{\mathcal I} \bar {\mathcal I}]}
\left( \frac{2S_I}{k} \right)^2 + \dots \right]
 \label{fluc-dis2}
\end{eqnarray}
 which can be checked against the result obtained via  Bender-Wu
recursion relations, an independent method to calculate
(\ref{fluc-dis2}).

 
 This implies that both in DW and SG cases,  not only does the leading term cancel, but also the sub-leading terms are canceled once we include the prefactor. Once again, there is a precise correspondence between the coefficients of the behavior of high orders of perturbation theory about the vacuum and the coefficients of the low orders of fluctuations about the 2-instanton sector: this is ``resurgence'' at work.

\section{Generating NP-physics  from P-physics} 
\label{sec:explicit}

At first sight (and naively), there is no real difference between the ZJJ and uniform WKB 
approaches.  However, the latter approach reveals a simple and elegant relation between perturbative and non-perturbative physics that is not obvious in the former.

In ZJJ, one  computes the perturbative function $B(E, g^2)$ and the non-perturbative function $A(E, g^2)$, and imposes an exact quantization condition. Although calculation of  $B(E, g^2)$  is straightforward, the evaluation of $A(E, g^2)$ is more challenging. 

In uniform WKB approach,  one computes the perturbative function $E(B, g^2)$ and the prefactor function $A(B, g^2)$, and imposes a global boundary condition.  This reveals an extremely simple (but non-obvious) relation between the two functions $E(B, g^2)$ and $A(B, g^2)$:
\begin{eqnarray}
\frac{\partial E}{\partial B}=- \frac{g^2}{S}\left(2B+g^2 \frac{\partial A}{\partial g^2}\right)
\label{eq:general-magic}
\end{eqnarray}
where $S$ is the numerical coefficient of the instanton action in  $\xi\equiv e^{-S/g^2}/\sqrt{\pi g^2}$. 
 This relation was not observed in \cite{zjj}, because the relation is  not apparent when looking at the expansions of the functions $B(E, g^2)$ and $A(E, g^2)$.(Note that  $E_{ZJ}\to \frac{E}{2}$, and $g_{ZJ}\to g^2$. We used ZJJ in  \cite{Dunne:2013ada}.)

Eq.(\ref{eq:general-magic}) has a magical implication:  that all the information in the non-perturbative expression $A(B, g^2)$ is completely determined by the perturbative expression $E(B, g^2)$.  
Thus, the non-perturbative computation of  $A(B, g^2)$ is actually unnecessary!  
The overall factor $S$ appearing in the formula can also be deduced from the  large order growth of $E(B, g^2)$, or  be computed trivially by usual instanton methods, but crucial thing is that it is also already encoded in late non-alternating terms  of the, for example,  ground state perturbative expansion $E(B=0+ \frac{1}{2}, g^2) \sim \sum_n a_n g^{2n} $ where  $a_n \sim n!/(2S)^n$. 

This is astonishing, especially in light of the extremely complicated non-perturbative multi-instanton analysis required to compute $A(E, g^2)$ in \cite{zjj}.
All features of the non-perturbative sector are encoded in the perturbative sector, provided we know the perturbative expansion $E(B, g^2)$  as a function of both the coupling $g^2$ and the level number parameter $B$. 

This is an explicit realization of \'Ecalle's statement that all information about the trans-series is encoded in the perturbative sector. For this double-well potential, this fact was noticed previously, in a slightly different form,  in a beautiful paper by \'Alvarez \cite{alvarez}. Below we show that it is  more general \cite{Dunne:2013ada}.

\subsection{Double-Well Potential}

For sake of comparison, we recall the DW potential expressions:
\begin{eqnarray}
E_{\rm DW}(B, g^2)&=&2B-2g^2\left(3B^2+\frac{1}{4}\right)
-2g^4\left(17 B^3+\frac{19}{4}B\right)
-2g^6\left(\frac{375}{2}B^4+\frac{459}{4}B^2+\frac{131}{32}\right)\nonumber\\
&&
-2g^8\left(\frac{10689}{4}B^5+\frac{23405}{8}B^3+\frac{22709}{64}B\right)-\dots
\label{ebdw}
\end{eqnarray}
\begin{eqnarray}
A_{\rm DW}(B, g^2)&=&\frac{1}{3g^2}+g^2\left(17B^2+\frac{19}{12}\right)
+g^4\left(125 B^3+\frac{153}{4}B\right)
+g^6\left(\frac{17815}{12}B^4+\frac{23405}{24}B^2+\frac{22709}{576}\right)\nonumber\\
&&
+g^8\left(\frac{87549}{4}B^5+\frac{50715}{2}B^3+\frac{217663}{64}B\right)-\dots
\label{adw3}
\end{eqnarray}
Notice the similarities between terms in the expansion of $A_{\rm DW}(B, g^2)$ and $E_{\rm DW}(B, g^2)$. To make this completely explicit, compute:
\begin{eqnarray}
\frac{\partial E_{\rm DW}(B, g^2)}{\partial B}&=& 2-12 B g^2
- 2 g^4 \left(51 B^2+\frac{19}{4}\right)
-2 g^6\left(750 B^3+\frac{459 B}{2}\right)
\nonumber\\
&&
- 2g^8 \left(\frac{53445 B^4}{4}+\frac{70215 B^2}{8}+\frac{22709}{64}\right)-\dots
   \label{dw-dedb}
   \end{eqnarray}
   And, for comparison, compute:
   \begin{eqnarray}
 -6g^4 \frac{\partial A_{\rm SG}(B, g^2)}{\partial g^2}&=&2
 -6 g^4 \left(17 B^2+\frac{19}{12}\right)
 -18 g^6 \left(125 B^3+\frac{153}{4}B\right)
  \nonumber\\
  && 
  -18 g^8\left(\frac{17815}{12}B^4+\frac{23405}{24}B^2+\frac{22709}{576}\right) -\dots 
  \label{dw-dadg}
     \end{eqnarray}
   We deduce the remarkably simple relation between the perturbative expression $E_{\rm DW}(B, g^2)$ and $A_{\rm DW}(B, g^2)$:
\begin{eqnarray}
\frac{\partial E_{\rm DW}}{\partial B}=-12\, B\, g^2-6g^4\frac{\partial A_{\rm DW}}{\partial g^2}
\label{dw-magic}
\end{eqnarray}
which is nothing but (\ref{eq:general-magic}) with $S= 1/6$, the instanton action. 
This means that the non-perturbative expression $A_{\rm DW}(B, g^2)$ is completely determined by the perturbative expression $E_{\rm DW}(B, g^2)$.

\subsection{Sine-Gordon Potential}

Remarkably, exactly the same thing happens for the SG potential. 
Again, for the sake of comparison, we recall the expressions:
\begin{eqnarray}
E_{\rm SG}(B, g^2)&=& B-  \frac{g^2}{2}  \left(B^2+\frac{1}{4}\right)
- \frac{g^4}{8}  \left(B^3+\frac{3 B}{4}\right) 
- \frac{g^6}{32} \left(\frac{5   B^4}{2}+\frac{17 B^2}{4}+\frac{9}{32}\right) \nonumber\\
   &&-  \frac{g^8}{128}  \left(\frac{33 B^5}{4}+\frac{205 B^3}{8}+\frac{405 B}{64}\right) +\dots
\label{bsg2b}
\end{eqnarray}
\begin{eqnarray}
A_{\rm SG}(B, g^2)&=&\frac{4}{g^2}+\frac{g^2}{4}\left(3 B^2+\frac{3}{4}\right) 
+\frac{g^4}{16}\left(5 B^3+\frac{17 B}{4}\right)
 +\frac{5g^6}{4096} \left(176 B^4+328 B^2+27\right) \nonumber\\
   &&+\frac{9 g^8}{16384} \left(336 B^5+1120 B^3+327 B\right)    +\dots
 \label{asg2b}
\end{eqnarray}
%
Notice again the similarities between terms in the expansion of $A_{\rm SG}(B, g^2)$ and $E_{\rm SG}(B, g^2)$. To make this completely explicit, compute:
\begin{eqnarray}
\frac{\partial E_{\rm SG}(B, g^2)}{\partial B}&=& 2- B g^2
-\frac{g^4}{8} \left(3 B^2+\frac{3}{4}\right)
-\frac{g^6}{32}\left(10 B^3+\frac{17 B}{2}\right)
\nonumber\\
&&
-\frac{g^8}{128}\left(\frac{165 B^4}{4}+\frac{615 B^2}{8}+\frac{405}{64}\right)-\dots
   \label{sg-dedb}
   \end{eqnarray}
   And, for comparison, compute:
   \begin{eqnarray}
 -\frac{1}{2}g^4 \frac{\partial A_{\rm SG}(B, g^2)}{\partial g^2}&=&2
 -\frac{g^4}{8}\left(3 B^2+\frac{3}{4}\right)
 -\frac{g^6}{32}
  \left(10 B^3+\frac{17 B}{2}\right)
  \nonumber\\
  && 
  -\frac{g^8}{128}\left(\frac{165 B^4}{4}+\frac{615 B^2}{8}+\frac{405}{64}\right)-\dots 
  \label{sg-dadg}
     \end{eqnarray}
   We deduce the remarkably simple relation: 
\begin{eqnarray}
\frac{\partial E_{\rm SG}}{\partial B}=  - B\, g^2-\frac{1}{2}g^4\frac{\partial A_{\rm SG}}{\partial g^2}
 = - \frac{2 (g^2/4) B } {S} - \frac{(g^2/4)^2}{S}  \frac{\partial A}{\partial (g^2/4) }
\label{sg-magic}
\end{eqnarray}
In the second equality, we observe that instanton action $S=1/2$  and expansion parameter $g^2 \rightarrow g^2/4$, and hence, this is again same as  (\ref{eq:general-magic}) \cite{Dunne:2013ada}.

\subsection{Fokker-Planck Potential}

In \cite{zjj}, ZJJ present expressions for $B(E, g)$ and $A(E, g)$ for  the Fokker-Planck  potential (in this section we use their conventions for the coupling and normalizations):
\begin{eqnarray}
V_{\rm FP}(y)=\frac{1}{2}y^2(1-y)^2+g\left( y-\frac{1}{2}\right)
\label{fp}
\end{eqnarray}
This is essentially a double-well potential, with a linear symmetry breaking term. It can be thought of as the SUSY QM  version of the double-well problem.  ZJJ give the results:
\begin{eqnarray}
B_{\rm FP}(E, g)&=& E+3g E^2 +g^2\left(35 E^3+\frac{5}{2}E\right)+g^3\left(\frac{1155}{2}E^4+105E^2\right)+\dots
\end{eqnarray}
The non-perturbative function $A_{\rm FP}(E, g)$ is \cite{zjj}:
\begin{eqnarray}
A_{\rm FP}(E, g)&=& \frac{1}{3g}+g\left(17 E^2+\frac{5}{6}\right)+g^2\left(227 E^3 +\frac{55}{2}E\right)
\nonumber\\
&&
+g^3\left(\frac{47431}{12}E^4+\frac{11485}{12}E^2+\frac{1105}{72}\right)+\dots
\end{eqnarray}
Inverting, to write $E$ as a function of $B$, we find
\begin{eqnarray}
E_{\rm FP}(B, g)&=& B-3g B^2 -g^2\left(17 B^3+\frac{5}{2}B\right)-g^3\left(\frac{375}{2}B^4+\frac{165}{2}B^2\right)+\dots
\end{eqnarray}
Inserting the expression for $E=E_{\rm FP}(B, g)$ we obtain
\begin{eqnarray}
A_{\rm FP}(B, g)&=&
\frac{1}{3g}+g\left(17 B^2+\frac{5}{6}\right)+g^2\left(125 B^3 +\frac{55}{2}B\right)
\nonumber\\
&&
+\frac{5}{72} g^3\left(21378 B^4+11370 B^2+221\right)+\dots
\end{eqnarray}
 Thus, we observe the relation
   \begin{eqnarray}
   \frac{\partial E_{\rm FP}(B, g)}{\partial B}&=& -6B g-3  g^2  \frac{\partial A_{\rm FP}(B, g)}{\partial g}
   \end{eqnarray}
So, again, the non-perturbative function $A_{\rm  FP}(B, g)$ is determined by the perturbative function $E_{\rm FP}(B, g)$.

\subsection{Symmetric AHO Potential}
Another example studied by ZJJ is the $O(d)$ symmetric anharmonic oscillator, with potential (in this section we use their conventions for the coupling and normalizations):
\begin{eqnarray}
V_{\rm AHO}(\vec{x})=\frac{1}{2}\vec{x}^2+g(\vec{x}^2)^2
\label{aho}
\end{eqnarray}
The  radial problem with angular momentum $l$ leads to a spectral problem characterized by a parameter $j=l+d/2-1$. ZJJ find the following expressions,  as a function of $j$:
\begin{eqnarray}
B_{\rm AHO}(E, g)&=& E+\frac{g}{2} \left(j^2-3 E^2-1\right)
+\frac{g^2}{4}  \left(-15 j^2 E+35 E^3+25 E\right)\nonumber\\
&&+   
\frac{g^3}{16} \left(-35 j^4+630j^2 E^2+ 210 j^2-1155 E^4-1470 E^2-175\right)+\dots 
\end{eqnarray}
 The function $A_{\rm AHO}(E, g)$ is \cite{zjj} (note: we adopt the sign convention from \cite{zjj})
   \begin{eqnarray}
A_{\rm AHO}(E, g)&=& -\frac{1}{3 g} +
g \left(\frac{3  j^2}{4}-\frac{17 E^2}{4}-\frac{19}{12}\right)
+g^2 \left(-\frac{77 j^2 E}{8}+\frac{227 E^3}{8}+\frac{187 E}{8}\right)\nonumber\\
&& + g^3 \left(-\frac{341 j^4}{64}+\frac{3717 j^2 E^2}{32}+\frac{1281 j^2}{32}-\frac{47431 E^4}{192}-\frac{34121
   E^2}{96}-\frac{28829}{576}\right)+\dots
   \end{eqnarray}
Inverting, to write $E$ as a function of $B$, we find
\begin{eqnarray}
E_{\rm AHO}(B, g)&=& B+\frac{1}{2} g \left(3 B^2-j^2+1\right)+
\frac{1}{4} g^2 \left(-17 B^3+9 B j^2-19 B\right)\nonumber
\\
&&+\frac{1}{16} g^3 \left(375 B^4-258 B^2 j^2+918
   B^2+11 j^4-142 j^2+131\right)+\dots
   \end{eqnarray}
  Converting $A$ to a function of $B$, we find
  \begin{eqnarray}
A_{\rm AHO}(B, g)&=& -\frac{1}{3 g}+\frac{1}{12} g \left(-51 B^2+9 j^2-19\right)+\frac{1}{8} g^2 \left(125 B^3-43 B j^2+153 B\right)\nonumber\\
&&+\frac{1}{576} g^3
   \left(-53445 B^4+26730 B^2 j^2-140430 B^2-909 j^4+14778 j^2-22709\right)   +\dots
   \end{eqnarray}
   Thus, we see that for all $j$, we have the relation
   \begin{eqnarray}
   \frac{\partial E_{\rm AHO}(B, g)}{\partial B}&=& 3B g+3  g^2  \frac{\partial A_{\rm AHO}(B, g)}{\partial g}
   \end{eqnarray}
So, again, the non-perturbative function $A_{\rm AHO}(B, g)$ is determined by the perturbative function $E_{\rm AHO}(B, g)$.

\section{Conclusions}

In this paper we have given an elementary derivation, using a uniform WKB expansion, of the appearance of trans-series expressions of the form (\ref{trans})  for energy eigenvalues in quantum problems with degenerate harmonic minima. We have shown that this trans-series form is generic for such problems because it can be related, in the small $g^2$ limit, to basic  analyticity properties of the parabolic cylinder functions that underly harmonic vacuum problems. The global boundary conditions that relate neighboring vacua  for the double-well potential and the periodic Sine-Gordon potential problems lead naturally to resurgent relations connecting different parts of the trans-series expansion, again due to analyticity properties of the parabolic cylinder functions. 

The trans-series expansion unifies the perturbative and non-perturbative sectors, in such a way that ambiguities are cancelled between sectors, yielding  real and unambiguous results. The global boundary conditions are expressed in terms of two functions, the perturbative energy $E=E(B, g^2)$, and a non-perturbative function $A=A(B, g^2)$ that contains the single-instanton factor and fluctuations around it. Here $B=N+\frac{1}{2}$, where $N$ is an integer labeling the energy level or band. Given these two functions, the global  boundary condition generates the entire trans-series expansion, incorporating {\it all} multi-instanton effects to all orders both perturbatively and non-perturbatively. 

Finally, we have shown that there is a remarkably simple relation between the functions $E(B, g^2)$ and $A(B, g^2)$, which means that $A(B, g^2)$ is completely determined by knowledge of $E(B, g^2)$. 
{Thus, the entire trans-series, including all perturbative, non-perturbative and quasi-zero-mode terms, is encoded in the perturbative expansion  \cite{Dunne:2013ada}. In other words, the fluctuations around the vacuum saddle point contain information about all other non-perturbative saddles, including 
their non-perturbative actions as well as perturbative   fluctuations around them.
 This is  a physical manifestation of the mathematical concept of resurgence. A more complete understanding of this remarkable phenomenon in the language of path integrals would facilitate further application of the ideas and methods of resurgence  to quantum field theory and string theory  \cite{marino,Marino:2012zq,schiappa, Endres:2013sda, Krefl:2013bsa, Pius:2013tla, Chan:2013wya, Dunne:2012ae,Argyres:2012ka}.
\\

We acknowledge support from U.S. DOE grants DE-FG02-13ER41989 and
DE-FG02-92ER40716 (G.D.),  and DE-FG02-12ER41806 (M.\"U.). G.D. also thanks the CoEPP and CSSM, School of Chemistry and Physics, at the University of Adelaide, and the Physics Department at the Technion, for hospitality and support during the final stages of this work.

\end{document}